\documentclass[a4paper,12pt]{article}
\pdfoutput=1
\usepackage{jheppub}
\usepackage{epsfig}
\usepackage{amsmath,amssymb,amsfonts,verbatim,float,mathtools}
\usepackage[usenames,dvipsnames]{xcolor}

\usepackage{times}
\usepackage{booktabs}

\usepackage[english]{babel}

\def\F{{{\cal F}}}
\def\Real{{\rm Re}\,}

\newcommand{\rir}{r_\text{IR}}
\newcommand{\ruv}{r_\text{UV}}
\newcommand{\LIR}{L_\text{IR}}
\newcommand{\zuv}{z_\text{UV}}

 \newcommand{\rdw}{r_{\text{DW}}}
\newcommand{\huv}{h_{\text{UV}}}
\newcommand*\FF{\,_2F_1}
\newcommand*\FG[7]{\,_2F_1\left(\frac{#1}{#2},\frac{#3}{#4},\frac{#5}{#6},#7\right)}
\newcommand{\tuv}{t_{\text{UV}}}
\newcommand{\auv}{A_{\text{UV}}}
\newcommand{\frn}{f_{\text{RN}}}
\newcommand{\ep}{\epsilon}

\newcommand{\half}{{1\over2}}
\newcommand{\OO}{{\cal O}}

\newcommand{\Axz}{A_x^{(0)}}
\newcommand{\Axo}{A_x^{(1)}}

\title{\Large \bf Conductivity and entanglement entropy of high dimensional holographic superconductors}
\author{Antonio M. Garc\'ia-Garc\'ia and}
\author{Aurelio Romero-Berm\'udez}
\affiliation{TCM group, Cavendish Laboratory,  University of Cambridge,\\ J.J. Thomson Av., Cambridge, CB3 0HE, United Kingdom.}
\emailAdd{amg73@cam.ac.uk, abr31@cam.ac.uk}

\abstract{
We investigate the dependence of the conductivity and the entanglement entropy on the space-time dimensionality $d$ in two holographic superconductors: one dual to a quantum critical point with spontaneous symmetry breaking, and the other modeled by a charged scalar that condenses at a sufficiently low temperature in the presence of a Maxwell field. In both cases the gravity background is asymptotically Anti de Sitter (AdS). In the large $d$ limit we obtain explicit analytical results for the conductivity at zero temperature and the entanglement entropy by a $1/d$ expansion. We show that the entanglement entropy is always smaller in the broken phase. As dimensionality increases, the entanglement entropy decreases, the coherence peak in the conductivity becomes narrower and the ratio between the energy gap and the critical temperature decreases. These results suggest that the condensate interactions become weaker in high spatial dimensions.}
\begin{document}
\maketitle
%%%%%%%%%%%%%%%%%%%%%%%%%%%%%%%%%%%%
\section{Introduction}
%%%%%%%%%%%%%%%%%%%%%%%%%%%%%%%%%%%%
It is a well known fact in condensed matter and statistical physics that the dynamics of many systems simplifies drastically in the limit of large space dimensions $d-1$ \cite{Vollhardt1989,Kotliar1992,Monte1992,scalapino1994,sclarged1993,Fisher1964,oneoverd1995,Kotliarrev1996,percolation1976}. Analytical results are typically obtained for $d \to \infty$ \cite{Kotliar1992,Vollhardt1989} and, in some cases, it is also possible to  compute explicitly small corrections \cite{oneoverd1995} due to a large but finite dimensionality by a $1/d$ expansion. A typical example is the Hubbard model in the strong coupling region where in the large $d$ limit the problem maps onto a mean-field quantum impurity model that is solved self consistently. Meaningful results are only obtained \cite{Vollhardt1989} after the kinetic energy is properly rescaled so that the trivial non-interacting limit is avoided for $d \to \infty$. The application of these ideas to the Hubbard model was a key step for the later development of dynamical field theory \cite{Kotliarrev1996}. Another problem in which large $d$ expansion is relevant is that of a particle in a random potential. According to the selfconsistent theory of localization, \cite{abou1973} explicit analytical results for the critical disorder that induces a metal-insulator transition are only  known for a Cayley tree geometry which corresponds to a lattice of infinite dimensionality. However, there is still qualitative agreement with the numerical results in a three dimensional lattice \cite{cuevas2007}.

Similarly, many problems in percolation \cite{percolation1976} and spin chains \cite{Fisher1964} have explicit analytical results in the limit of large spatial dimensions. In many of these cases just keeping the leading term in the $1/d$ expansion is enough to find good agreement with experimental or numerical results \cite{Monte1992} in $d=3$. In the context of quantum gravity, large $d$ expansions have also been employed  \cite{strominger1982,bohr2004} to simplify Feynman diagrams in a spirit similar to the large $N$ approximation, broadly used in quantum chromodynamics, $N = 3$, and other gauge theories. However, renormalization of quantum theories of gravity is even more problematic as dimensionality increases so it is not clear whether it is a viable approximation scheme. The situation is different in classical theories of gravity which are finite for any dimensionality. The study of properties of black holes \cite{laflamme1993} and general relativity  \cite{Emparan2013} in large dimensions has shown that there are intriguing features that only occur for a sufficiently large number of dimensions.
More recently \cite{Emparan2014,Emparan2014a,Emparan2015} this large $d$ limit was studied in the context of AdS spaces and then applied, by AdS/CFT techniques, to the study of holographic superconductors \cite{Emparan2014}.

One of the main conclusions of \cite{Emparan2014} is that it is possible to find an explicit analytical expression of the critical temperature in the limit of large dimensionality and negligible backreaction of the scalar on the metric and on the gauge field. Even for $d=2+1$, this simple analytical prediction for the critical temperature is already a good approximation of the numerical results. Moreover, as dimensionality increases the condensation of the scalar occurs always close to the horizon as the gravitational effects of the black hole are only important in this region.

In this paper we continue the study of holographic superconductors in the large $d$ limit with a twofold motivation. Firstly, we aim to emphasize the usefulness of large $d$ expansions in holography by carrying out analytical calculations of the entanglement entropy and the conductivity that are only possible in this limit. Secondly, we seek to clarify the qualitative effect of dimensionality in holography. We have found that as $d$ increases the coherence peak becomes narrower and the ratio between the energy to break the condensate and the critical temperature decreases. This is a strong suggestion that the effective coupling that controls the interactions of the condensate seems to be weaker as dimensionality increases. 

The organization of the paper is as follows: In section \ref{models} we introduced the two models that we employ to study a large $d$ holographic superconductor, then in section \ref{conT} we compute numerically the conductivity up to $d=9$. Based on these results we compute in section \ref{order} the superconducting energy gap, roughly the maximum of the conductivity, and the order parameter $\langle\OO\rangle$ as a function of $d$. We also discuss certain ambiguity in the relation between these two quantities. In section \ref{Sec:cond} we study analytically at $T=0$, the low and large frequency-dependence of the electrical conductivity. Similarly, in section \ref{Sec:EE}, we provide simple analytical expressions for the entanglement entropy between a rectangular strip and its complement in the boundary; we analyze both the case $T=0$ and $T\sim T_c$.

%%%%%%%%%%%%%%%%%%%%%%%%%%%%%%%%%%%
%%%%%%%%%%%%%%%%%%%%%%%%%%%%%%%%%%%
\section{Models}\label{models}
%%%%%%%%%%%%%%%%%%%%%%%%%%%%%%%%%%%
We study the dimensional dependence of holographic  superconductivity \cite{Gubser2008,Hartnoll2008}  in two models, one at $T=0$ and other at $T>0$. 
%%%%%%%%%%%%%%%%%%%%%%%%%%%%%%%%%%%%
\subsection{d-dimensional holographic superconductivity at $T=0$}\label{T0gubser}
%%%%%%%%%%%%%%%%%%%%%%%%%%%%%%%%%%%%
For the $T=0$ limit we choose the model introduced in Ref.\cite{Gubser2009}, to describe a quantum critical point with spontaneous symmetry breaking,
\begin{equation}\label{action}
\begin{split}
S=&\int d^{d+1}x \sqrt{-g} \left[R-\frac{1}{4}F^{2}-|\partial_{\mu}\psi-iq^{2}A_{\mu}\psi|^{2}-V(|\psi|)\right]\hspace{-1mm},
\end{split}
\end{equation}
where 
\begin{equation}\label{eq:V}
V(|\psi|)=2\Lambda + m^{2}|\psi|^{2}+\frac{u}{2}|\psi|^{4},
\end{equation}
 $\Lambda=-d(d-1)/2L^{2}$ is the cosmological constant,  $m^2 < 0$ is the scalar mass and $u>0$. Symmetry breaking is directly related to the existence of a minimum of the potential at $|\psi|=\psi_{\text{IR}}=\sqrt{-m^{2}/u} \neq 0$. 

Following \cite{Gubser2009} we consider the metric ansatz
\begin{equation}\label{metric}
ds^{2}=e^{2A(r)}\left(-h(r)dt^{2}+dx^idx^i\right)+\frac{dr^{2}}{h(r)},
\end{equation}
$i=1,\dots,d-1$, such that in the infrared limit $A(r)=r/L_{\text{IR}}$ and $h(r)=1$ where $L_{\text{IR}}$ is defined through ${-d(d-1)\over \LIR^2}\equiv V(|\psi_{\text{IR}}|)$.

In order to recover the SO($d$-1,1) Lorentz symmetry and SO($d$,2) conformal symmetry deep in the IR  the metric should approach   
\begin{equation}\label{metricIR}
ds^{2}_{\text{IR}}=e^{2r/L_{\text{IR}}}\left(-dt^{2}+dx^idx^i\right)+dr^{2}
\end{equation}
where we have imposed
\begin{equation}\label{hIR}
h(r)\to h_{\text{IR}}=1,\ A(r)\to {r\over \LIR},\ \mbox{as } r\to-\infty.
\end{equation}

Similarly, in the UV limit, the appropriate symmetries are restored provided,
\begin{equation}\label{hUV}
h(r)\to h_{\text{UV}},\ A(r)\to A_{\text{UV}} {r\over L},\ \mbox{as } r\to\infty,
\end{equation}
 with $h_{\text{UV}}$ and $ A_{\text{UV}} $ constants related by  the $rr$ component of the Einstein equations:
\begin{equation}\label{constraint}
\begin{split}
&(d-1)(A'h'h+dh^2A'^2)+hV(|\psi|)-h^2\psi'^2-e^{-2A}q^2\psi^2\phi^2+{h\over2}e^{-2A}\phi'^2=0,
\end{split}
\end{equation}
where $\phi$ is the $t$ component of the gauge field. Evaluated at the UV boundary, the previous equation, yields
\begin{equation}\label{relation}
\huv={1\over\auv^2}.
\end{equation}
Moreover, the null energy condition requires $\huv>h_{\text{IR}}=1$ \cite{Gubser2009} which means that $\auv<1$. At the same time, $A(r)$ must increase monotonically in the whole range $-\infty<r<\infty$ and the slope in the UV-limit must be lower than in the IR-limit, i.e., $\auv/L<1/\LIR$, \cite{Freedman1999}.

The resulting equations of motion are,
\begin{equation}
\begin{split}
&\psi''+\psi' \left({h'\over h}+ d A'\right)+\psi {q^2\phi^2 \over e^{2A}h^2}+{V'(|\psi|)\over 2h}=0,\ \ \phi''+\phi'(d-2)A'-\phi{2\psi^2q^2\over h}=0,\\
&\hspace{5mm}h''+dh'A'-{2\over h}q^2\phi^2\psi^2 e^{-2A}-{\phi'}^2e^{-2A}=0,\ \ A''+{1\over d-1}{\psi'}^2+{e^{-2A}q^2\phi^2\psi^2\over (d-1)h^2}=0\\
\end{split}
\end{equation}
with boundary conditions in the IR-limit ($r\to-\infty$),
\begin{equation}\label{bcIR}
\phi\sim\phi_0 e^{{r\over L_{\text{IR}}}[\Delta_{\phi_\text{IR}}-(d-2)]},\ \ \psi=\psi_{\text{IR}}+a_\psi e^{{r\over L_{IR}}(\Delta_{\psi_\text{IR}}-d)},\\
\end{equation}
where $\Delta_{\phi_\text{IR}}$ and $\Delta_{\psi_\text{IR}}$ are the larger roots of: $\Delta_{\phi_\text{IR}}[\Delta_{\phi_\text{IR}}-(d-2)]=2q^2\psi_\text{IR}^2\LIR^2$ and $\Delta_{\psi_\text{IR}}(\Delta_{\psi_\text{IR}}-d)={1\over2}V''(\psi_\text{IR})\LIR^2=-2m^2L_{\text{IR}}^2$.
Similarly in the UV limit ($r\to\infty$),
\begin{equation}\label{bcUV}
\phi=\mu-\rho e^{-\Delta_{\phi_\text{UV}}{r\over  L}},\ \ \psi=\psi_\text{UV} e^{-{r\over L}\auv(d-\Delta_{\psi_\text{UV}})},
\end{equation}
where, $\Delta_{\phi_\text{UV}}=d-2$ and $\Delta_{\psi_\text{UV}}$ is the smaller root of: $\Delta_{\psi_\text{UV}}(\Delta_{\psi_\text{UV}}-d)=m^2L^2/(h_{\text{UV}}\auv^2)$. The boundary conditions for $h$ and $A$ are given in eqs. (\ref{hIR}) and (\ref{hUV}). Moreover, we will take the parameters $m^2$ and $u$ such that the operators dual to $\psi$ and $\phi$ are irrelevant in the IR so that the IR AdS space is a fixed point of the RG flow. Repeating the argument presented in \cite{Gubser2009GS} it is straightforward to see this corresponds, in our notation, to:
\begin{equation}
\Delta_{\psi_\text{IR}}>d,\ \ \Delta_{\phi_\text{IR}}> d-1.
\end{equation}

%%%%%%%%%%%%%%%%%%%%%%%%%%%%%%%%%%%%
\subsection{d-dimensional holographic superconductivity at $T>0$}\label{Sec:larger0}
%%%%%%%%%%%%%%%%%%%%%%%%%%%%%%%%%%%%
For the study of holographic superconductors at finite temperature we employ the, by now, standard model introduced in \cite{Gubser2008,Hartnoll2008} by coupling anti-de Sitter gravity to a Maxwell field and a charged scalar and a quadratic (in $|\psi|$) potential. Here we state the action and equations of motion in $d$ dimensions directly in order to settle down notation and refer to the reviews Refs. \cite{Herzog2009,Faulkner2010} for more details.
The action is given by, 
\begin{equation}\label{action2}
\begin{split}
S=&\int d^{d+1}x \sqrt{-g} \left[R-\frac{1}{4}F^{2}-{|D_\mu\psi|^{2}}-V(|\psi|)\right],\ \ V(|\psi|)=-2\Lambda + m^{2}|\psi|^{2},
\end{split}
\end{equation}
with $D_\mu=\partial_{\mu}-iq^{2}A_{\mu}$. In probe limit, corresponding to a negligible backreaction of the scalar and the Maxwell field on the geometry, is simply given by the planar-Schwarzchild AdS black hole,
\begin{equation}
ds^2=-{r^2\over L^2}h(r)dt^2+{L^2dr^2\over r^2h(r)}+r^2dx^idx^i,\ \ i=1,\dots,d-1,
\end{equation}
with $h(r)=1-r_0^d/r^d$. Assuming for the moment that the only component of the Maxwell field is $A_t=\phi(r)$ it is straightforward to obtain:
\begin{equation}\label{eomnum}
\begin{split}
&\psi''+\psi' \left({h'\over h}+ {d+1\over r}\right)+\psi {\phi^2 \over r^4h^2}+{V'(|\psi|)\over 2r^2h}=0,\ \ \phi''+\phi'{d-1\over r}-\phi{2\psi^2\over r^2 h}=0.\\
\end{split}
\end{equation}
The boundary conditions are fixed from to the usual expansions:
\begin{equation}\label{bdy_cond}
\begin{split}
&\hspace{6mm}\psi(r\to\infty)={\alpha\over r^{\Delta_-}}+{\beta\over r^{\Delta_+}}+\dots,\ \ \phi(r\to\infty)=\mu+{\rho\over r^{d-2}}+\dots\\
&\psi(r\to r_0)=\psi_0+\psi_1\left(1-{r_0\over r}\right)\dots,\ \ \phi(r\to r_0)=\phi_1\left(1-{r_0\over r}\right)+\dots,\\
\end{split}
\end{equation}
where $\Delta_{\pm}=\half \left(d\pm\sqrt{d^2+4m^2L^2}\right)$ and $\psi_1$ is given in terms of the undetermined constants $\psi_0$, $\phi_1$.

%%%%%%%%%%%%%%%%%%%%%%%%%%%%%%%%%%%%
%%%%%%%%%%%%%%%%%%%%%%%%%%%%%%%%%%%%
\section{Electrical conductivity in the large $\boldsymbol d$ limit: $\boldsymbol{T>0}$ case}\label{conT}
%%%%%%%%%%%%%%%%%%%%%%%%%%%%%%%%%%%%
We start our analysis by computing the electrical conductivity, $\sigma$, at $T>0$. For the sake of completeness we review the procedure to compute it for a general $d$. To this end one should add a perturbation to the vector potential $\delta A=A_x$ as well as one to the metric $\delta g=g_{tx}$. However, we will solve for $\sigma$ numerically in the probe limit where,
\begin{equation}\label{metric:probe}
ds^2={1\over z^2}\left(f(z) dt^2+{1\over f(z)}dz^2+dx_i^2\right), \ \ f(z)=1-z^d, \ \ i=1,\dots,d-1,
\end{equation}
with $z=1/r$, and choosing the horizon position $z_0=1$ and $L=1$. As usual, the linear response of an operator, in our case the current $J^\mu(x)$, to an external source or field perturbation, $A_x$, is related, in momentum space, to the retarded Green's function, \cite{Herzog2009}:
\begin{equation}
\delta J^x(k)=\tilde G^{xx}_R(k)\delta\tilde  A_x(k),
\end{equation}
where $k=(\omega,\vec{k})$ is the $d$-momentum and $\tilde G^{xx}_R(k)$ is the Fourier transform of the retarded Green's function. Moreover, the charge current response to an electric field is $J^i(\omega)=\sigma^{ij}(\omega)E_j(\omega)$, with $E_x=-\partial_tA_x(t,z,x)$, $A_x(t,z,x)=e^{i\omega t}A_x(z,x)$, therefore, it follows that,
\begin{equation}\label{sigma:def}
\sigma^{xx}(\omega)={\tilde G_R^{xx}(\omega,0)\over  i \omega}.
\end{equation}
We now compute this Green's function following the procedure first outlined in Ref. \cite{Son2002}.  First, we write the Fourier transform of the vector potential,
\begin{equation}\label{AmuFT}
A_\mu(z,x)=\int {d^dk\over (2\pi)^d}e^{ikx}\tilde A_\mu(z,k)
\end{equation}
where $kx=-\omega t+\vec{k} \cdot \vec{x}$ and $\tilde A_\mu^{(0)}(k)= \tilde A_\mu(z=0,k)$ is defined from the boundary value $A_\mu(z=0,x)$. The Fourier transform of the gauge-field-part of the action leads to,
\begin{equation}\label{sgauge}
\begin{split}
&S_{gauge}=\left.\int {d^dk\over (2\pi)^d}\F(k,z)\right|_{z=0}^{z=1}\hspace{-3mm}+\dots,\ \F(k,z)=-{\sqrt{-g}g^{zz}g^{xx}\over 2}\tilde A_x(z,-k)\partial_z \tilde A_x(z,k),\\
\end{split}
\end{equation}
where the dots correspond to terms not containing $A_x$ and its derivatives. The final expression for the conductivity is obtained by combining the proposal of Ref. \cite{Son2002} 
$\tilde G_R^{xx}(\omega,0)=-2{\delta^2\over\delta \tilde A_x^{(0)}(-k)\delta \tilde A_x^{(0)}(k)}\lim_{z\to0} \F(k,z)$ together with eq. \ref{sigma:def}, 
\begin{equation}\label{conductivity}
\mbox{Re}[\sigma(\omega)]=\left.{1\over i \omega}{\delta^2\over\delta \tilde A_x^{(0)}(-\omega)\delta \tilde A_x^{(0)}(\omega)}\lim_{z\to0}\sqrt{-g}g^{zz}g^{xx}\tilde A_x(z,-k)\partial_z \tilde A_x(z,k)\right|_{\vec k=0}\hspace{-3mm}.
\end{equation}
In order to compute $\tilde A_x(z,k)$ we write the equation for $A_x(z,x)$ in the fixed background given in eq. (\ref{metric:probe}). As was mentioned above, we assume a harmonic time dependence for $A_x$. The derivatives are taken with respect to the holographic coordinate, $z$:
\begin{equation}\label{eomAxprobe}
A_x''+\left({-d+3\over z}+{f'\over  f}\right)A_x'+\left({\omega^2\over f^2}-{|\vec k|^2\over z^2 f}-{2\psi^2\over z^2f}\right)A_x=0, \ \ f=1-z^d.
\end{equation}
Finally, we impose the usual boundary conditions, in-falling close to the horizon, $A_x\sim (f/z^2)^{-i\omega/d}(1+\dots)$ and $A_x\sim A_{x}^{(0)}+A_{x}^{(1)}g_d(z,\omega)$ close to the boundary. The function $g_d(z,\omega)$ is easily obtained, for each $d$, by solving the asymptotic expansion of eq. (\ref{eomAxprobe}), App. \ref{App:cond}. By combining  eqs. (\ref{AmuFT}) and (\ref{eomAxprobe}) we obtain an analytically solvable differential equation for $\tilde A_x(z,k)$ at zero spatial momentum, which in the $z \to 0$ limit reduces to,
\begin{equation}
\tilde A_x''(z,\omega)+\tilde A_x'(z,\omega){3-d\over z}+\tilde A_x(z,\omega) \omega^2=0.
\end{equation}
 We choose the regular solution at $z=0$. We note that at $\omega=0$ the conductivity $\mbox{Re}(\sigma)$ develops a delta function as a consequence of the translational invariance of the system. 

For odd $d$ we have now all the ingredients to compute the conductivity $\sigma(\omega)$ (\ref{conductivity}). However for even $d$, logarithmic divergences at non-zero $\omega$ appear \cite{Horowitz2008}. In order to study the large-$d$ limit of $\sigma$, it is enough to restrict our analysis to odd $d$. Therefore, in order to avoid the intricacies of adding the counterterms to the action to remove the divergences mentioned above, we take the prescription for $d=4$ given in \cite{Horowitz2008} and for $d=3,5,7,9$ we employ eq. (\ref{conductivity}). 

%%%%%%%%%%%%%%%%%%%%%%%%%%%%%%%%%%%%
\subsection{Numerical calculation of the conductivity at low temperature for $d \leq 9$}\label{numerics}
\begin{figure}[H]
\center
\includegraphics[scale=0.68]{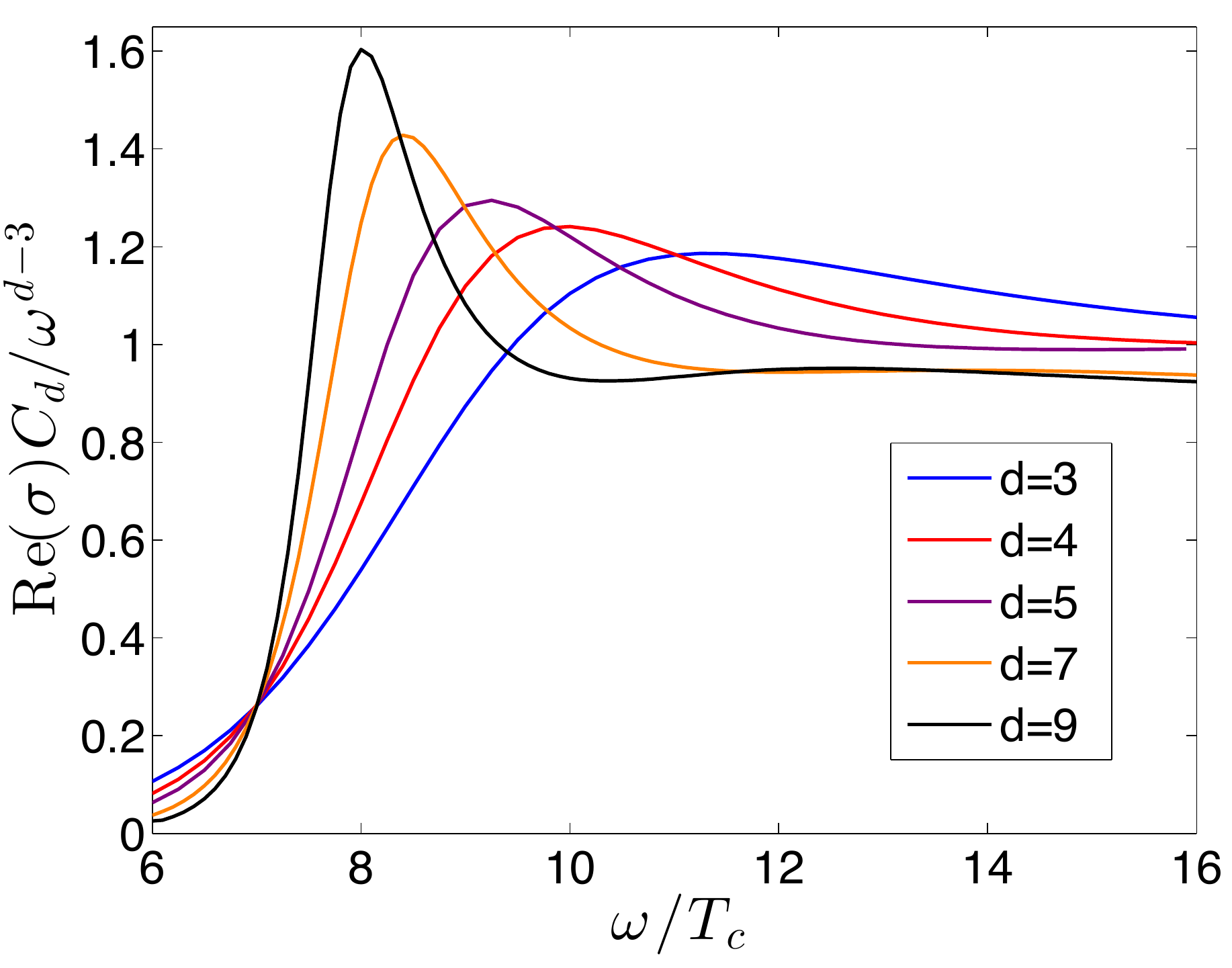}
\caption{Conductivity (\ref{conductivity}) in different dimensions for a massless scalar field at $T/T_c\sim 0.1$. As the dimensionality increases the coherence peak is narrower and moves to the region of lower frequencies.}
\label{condallomoTc}
\end{figure}

\begin{figure}[H]
\center
\includegraphics[scale=0.68]{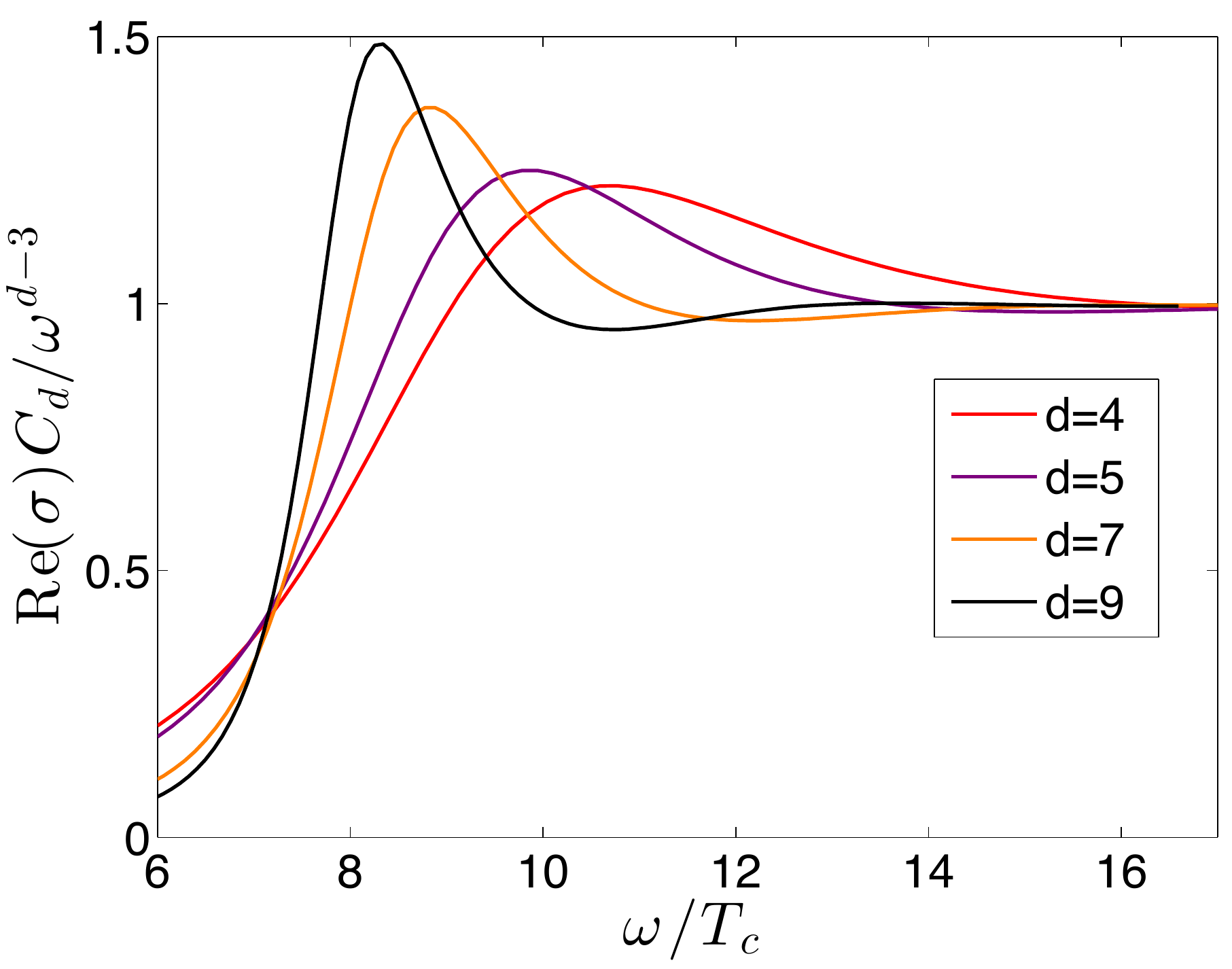}
\vspace{-0.4cm}
\caption{Conductivity (\ref{conductivity}) in different dimensions for $m^2L^2=d+1$ at $T/T_c\sim 0.6$. Results are similar to those of figure \ref{condallomoTc} for $m=0$}
\label{cond_other}
\end{figure}

\begin{figure}[H]
\center
\includegraphics[scale=0.6]{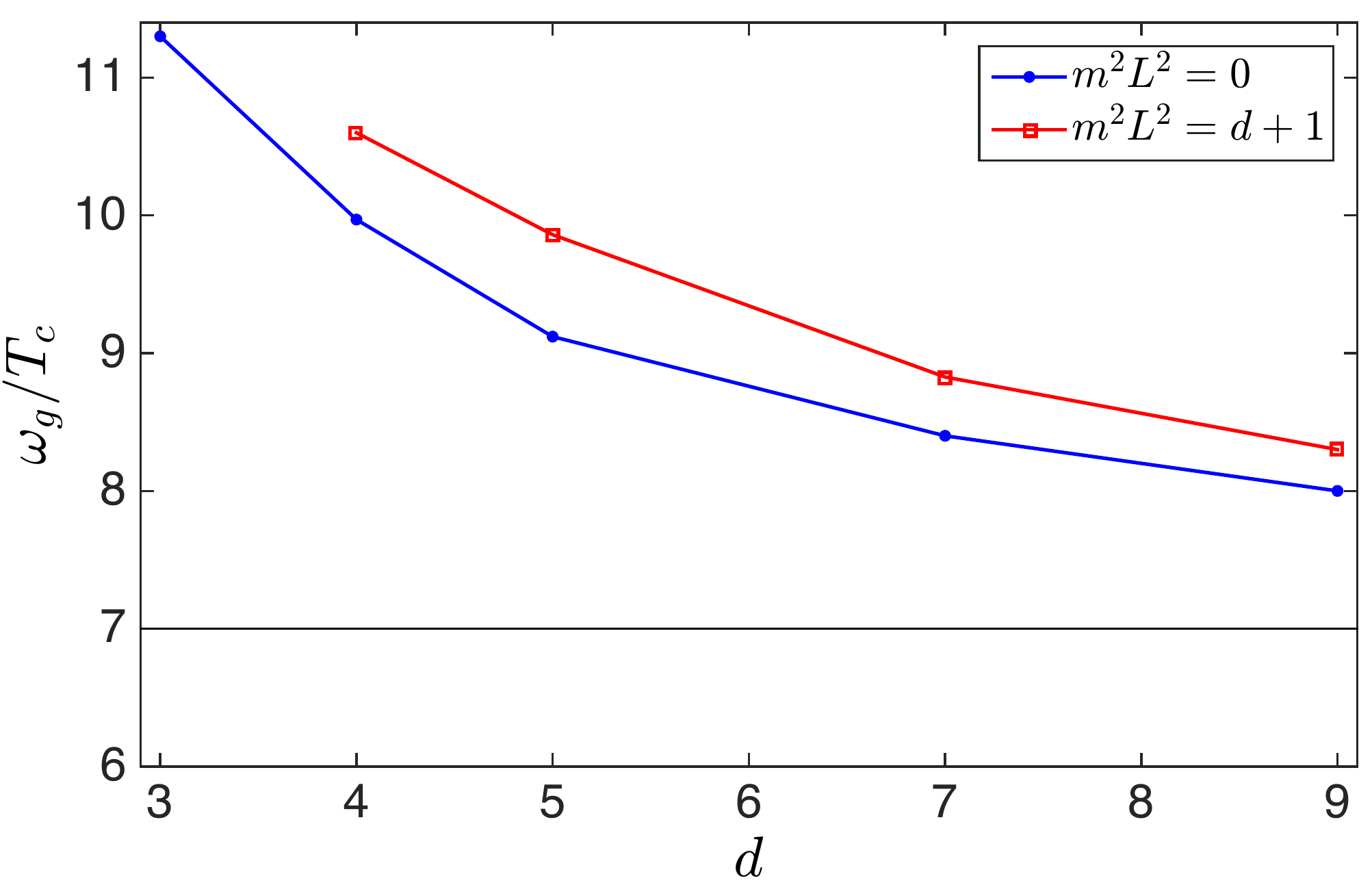}
\vspace{-0.5cm}
\caption{Ratio between the peak of the conductivity $\omega_g$ and the critical temperature for $m^2L^2=0$ ($\Delta=d$) at $T\sim0.1T_c$,  and $m^2L^2=d+1$ ($\Delta=d+1$) at $T\sim0.6T_c$ as a function of the dimensionality. It always decreases as $d$ increases and only depends weakly on the scalar mass $m^2$, see figures  \ref{condallomoTc}, \ref{cond_other}. The horizontal black line indicates the approximate position of the crossing point in figures  \ref{condallomoTc} and \ref{cond_other}, which we also expect to correspond to the location of the peak of the conductivity for $d\to \infty$ limit. We note that even in the $d \to \infty$ limit the ratio is still substantially larger than the BCS prediction $~3.528$.}
\label{OgTc}
\end{figure}

%%%%%%%%%%%%%%%%%%%%%%%%%%%%%%%%%%%%
In this section we compute numerically the electrical conductivity in the probe limit for $3, 4, 5, 7$ and $9$ dimensions of the dual boundary theory and for two scalar masses $m^2 = 0, \ d+1$.  We follow the procedure described in the previous section and solve the resulting differential equations by the shooting method.  See Appendix \ref{App:cond} for the specific expressions of the electrical conductivity in each dimension. The results depicted in figure \ref{condallomoTc} and figure \ref{cond_other} indicate that as dimensionality increases, the coherence peak becomes narrower and the position to the peak $\omega_g$ moves to lower frequencies. The physical interpretation of these features is clear. The condensate becomes less coupled as it costs less energy to break it (smaller $\omega_g$) as $d$ increases. Moreover, the effective bulk coupling also decreases as a narrower coherence peak is a signature of a longer life-time of the relevant excitations around $\omega_g$. A tentative explanation of this behaviour in the gravity dual is that \cite{Emparan2013,Emparan2014} as the dimensionality increases the condensation of the scalar gradually occurs closer to the horizon which corresponds to the less strongly interacting limit of the dual field theory. 
A natural question to ask is whether the gravity dual has a well defined limit for $d \to \infty$. In order to answer this question in figure \ref{OgTc} we plot $\omega_g/T_c$ as a function of $d$. The ratio decreases monotonically as $d$ increases and it is likely to converge to a finite value in the $d \to \infty$ limit still above the prediction $\sim 3.528$ of the Bardeen-Cooper-Schrieffer (BCS) theory of weakly coupled superconductors. It seems that this limiting value only depends weakly on the scalar conformal weight. More specifically we expect this result to hold provided that both the chemical potential, related to the kinetic energy, and the conformal weight, related to interaction energy, have the same scaling with $d$. It would be interesting to explore whether there exists a strict minimum bound for this quantity in the large $d$ limit. 
\vspace{-1cm}
A few comments are in order: a) we omit the case $d=3$ for $m^2L^2=d+1$ in what follows since we have observed an anomalous behavior of the AC conductivity similar to that reported in Ref.\cite{Horowitz2008}, b) in figure \ref{cond_other}, corresponding to $m^2L^2=d+1$, the `crossing point' where all curves meet, $\omega/T_c \sim 7$, is slightly blurred due to the presence of extra poles, not shown, at lower frequency \cite{Horowitz2008}, c) although for $m^2L^2=d+1$ we found difficult to decrease the temperature below $T/T_c\sim0.6$ it is clear, see figure \ref{OgTc}, that the behavior for both masses is strikingly similar.

%%%%%%%%%%%%%%%%%%%%%%%%%%%%%%%%%%%%
%%%%%%%%%%%%%%%%%%%%%%%%%%%%%%%%%%%%
\section{Relation between the order parameter $\boldsymbol{\langle \OO\rangle}$ and $\boldsymbol{\omega_g}$ in the large $\boldsymbol d$ limit}\label{order}
\begin{figure}[H]
\includegraphics[scale=0.7]{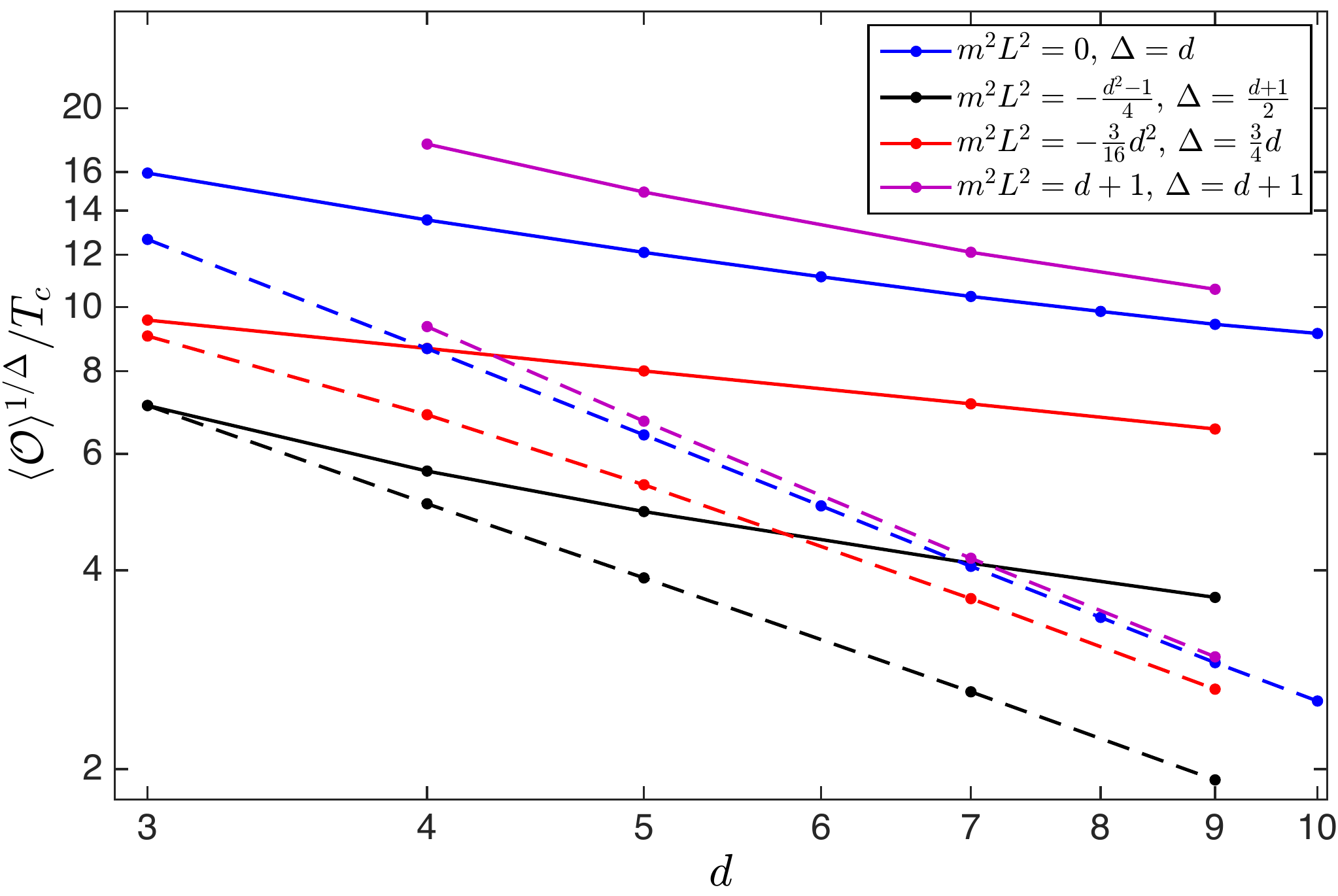}
\vspace{-4mm}
\caption{Ratio between the order parameter close to zero temperature and the critical temperature for different dimensions and scalar masses. Dashed lines correspond to the usual definition:  $\langle \OO\rangle = (2\Delta-d)\beta$, continuous lines include a speculative factor $\Gamma(\Delta)$, $\langle \tilde \OO\rangle =(2\Delta-d)\Gamma(\Delta)\beta$. Only in the latter case convergence of the ratio to a non-zero value in the $d \to \infty$ limit is likely.} 
\label{OoTc}
\end{figure}

%%%%%%%%%%%%%%%%%%%%%%%%%%%%%%%%%%%%
In BCS superconductors the coherence peak in the conductivity is simply two times the value of the order parameter also referred to as the superconducting energy gap. Physically it means that since a Cooper pair is composed of two electrons it takes twice the energy gap to break a Cooper pair and place these two electrons in the first state available above the Fermi energy. For strongly coupled superconductors there is no clear relation between these two quantities as the coherence peak broadens substantially and in some materials the quasiparticle picture based on the Fermi liquid approximation breaks down. However, in the context of holographic superconductivity it is well known \cite{Horowitz2009} that these two observables are still comparable, though the relation between them is not universal and different from the BCS prediction \cite{Basu2011}. We now study to what extent this relation still holds in the large $d$ limit. The order parameter $\langle \OO\rangle$ is computed by following the usual steps. First we find the  numerical solution of the equations of motion eq. (\ref{eomnum}) by the shooting method for a scalar field, $\psi(r)$, charged under the gauge field $A_t=\phi(r)$ in a non-dynamical Schwarzchild background, i.e., in the probe limit. We consider a fixed charge density and different scalar masses. The order parameter is simply $\langle \OO\rangle^{1\over\Delta}=\left[(2\Delta-d)\beta\right]^{1\over\Delta}$, where $\Delta$ is the conformal dimension of the operator dual to $\psi$, $d$ is the number of dimensions of the dual theory, and $\beta$ is  given in the boundary condition, eq. (\ref{bdy_cond}).
In Table \ref{table_omegag} we present results for $\omega_g$ and $\langle \OO\rangle$ for different dimensions and masses. As dimensionality increases $\langle \OO\rangle$ becomes much smaller than $\omega_g$. Indeed, it seems that the ratio $\langle \OO\rangle^{1/\Delta}/\omega_g \to 0$ as $d \to \infty$. Presently we do not have a solid explanation for this discrepancy. A finite value of the order parameter $\langle \OO\rangle$ in holographic superconductivity is interpreted as a signature of  spontaneously symmetry breaking rather than a energy gap in the spectrum. It might therefore be that these two quantities are not related and the similar value in low dimensions is a coincidence. Another more speculative explanation is that the standard recipe to compute $\langle \OO\rangle$
misses some dimensionality prefactor. We went over the original derivation of the expression for the order parameter but we could not find any discrepancy with the expression used above. However we found that by rescaling, see figure \ref{OoTc}, $\langle \OO\rangle$ by $\Gamma(\Delta)$ the ratio seems to converge to a finite positive value in the $d \to \infty $ limit.\footnote{Our numerical results suggest $[(2\Delta-d)\beta]^{1\over\Delta}\to$ constant for $d\to\infty$. Thus, a factor depending only on $d$ such as $\Gamma(d)$, instead of $\Delta$, does not result in a finite $\langle\OO\rangle^{1/\Delta}/T_c$ in the the limit $d \to \infty$.}  Whether this is just a coincidence or has a deeper physical meaning remains to be understood. Finally, we note the fact that the rescaling by $\Gamma(\Delta)$ depends on the scalar mass indicates that it is not related to the dimensional dependence of the coupling constant in the action which is usually set to the unity.
\begin{table}[H]
\begin{center}
\def\arraystretch{1.0}
\begin{tabular}{|l|c|c|c|c|c|}
\cline{2-6}
\multicolumn{1}{c|}{} &\rule{0pt}{16pt}$\omega_g\over T_c$&$\langle {\cal O}\rangle^{1/\Delta}\over T_c$&$\langle {\cal \tilde O}\rangle^{1/\Delta}\over T_c$&$\langle {\cal O}\rangle^{1/\Delta}\over \omega_g$&$\langle {\cal \tilde O}\rangle^{1/\Delta}\over \omega_g$\\
\hline
$d=3$&11.3&12.7&16.0&1.1&1.4\\
\hline
$d=4$&10.0&8.7&13.6&0.9&1.4\\
\hline
$d=5$&9.1&6.4&12.1&0.7&1.3\\
\hline
$d=7$&8.4&4.1&10.5&0.5&1.3\\
\hline
$d=9$&8.0&2.9&9.4&0.4&1.2\\
\hline
\end{tabular}
\caption{Comparison of the position of the conductivity (\ref{conductivity}) coherence peak with the order parameter ${\langle \OO\rangle}=(2\Delta-d)\beta$ and the alternate definition  ${\langle \tilde\OO \rangle}=(2\Delta-d)\Gamma(\Delta)\beta$, for $m^2L^2=0$ ($\Delta=d$).  Convergence for large $d$ is only observed after the order parameter is rescaled by $\Gamma(\Delta)$. We do not have a clear understanding of why the order parameter and $\omega_g$ have a different parametric dependence on the dimensionality.}
\label{table_omegag}
\end{center}
\end{table}

%%%%%%%%%%%%%%%%%%%%%%%%%%%%%%%%%%%%
%%%%%%%%%%%%%%%%%%%%%%%%%%%
\section{Analytical calculation of the conductivity at $\boldsymbol{T=0}$ for different dimensions}\label{Sec:cond}
%%%%%%%%%%%%%%%%%%%%%%%%%%%
We now switch to the background introduced previously in section\ref{T0gubser} to describe holographic superconductivity at $T=0$.
From now the main focus of the paper will be to compute analylitically  the conductivity and later the entanglement entropy in the large $d$ limit in order to illustrate the interest of large $1/d$ expansion in holography.

In this section we compute the electrical conductivity at zero temperature. As was mentioned previously one must consider fluctuations of $A_{x}(t,r)$ and $g_{tx}(t,r)$, \cite{Hartnoll2008}, which source an electric field $E_{x}$ and carries momentum $T_{tx}$. These perturbations are usually assumed to have a harmonic time dependence, $A_x(r)e^{-i\omega t}$, $g_{tx}(r)e^{-i\omega t}$. Furthermore, the Einstein and Maxwell equations are expanded in $g_{tx}(r)$ keeping only linear terms in $A_x(r)$,
\begin{equation}\label{eomAx}
\begin{split}
A_{x}''(r)&+A_{x}'(r)\left[(d-2)A'+\frac{h'}{h}\right]+\frac{A_{x}(r)}{h}\left[\frac{\omega^{2}}{he^{2A}}-2q^{2}\psi^{2}-\phi'^{2}e^{-2A}\right]=0. \\
\end{split}
\end{equation}
We then impose that near the UV boundary $A_x(r)=A_0+A_1e^{-(d-2)A(r)}$. In the infra-red limit we expect the perturbation $A_x$ to become small. This is indeed the case for $d=3$ but not for $d \geq 4$ \cite{zeng2013} where it grows exponentially for $r \to -\infty$. This cast doubts about the stability of the background to small perturbations in large dimensions. Indeed, it has been observed that the addition of a gauge field increases the temperature of the dual field theory \cite{tarrio2011} even in the limit of an extremal black hole.  However a full stability analysis is beyond the scope of the paper as the main motivation here is to employ the large $d$ limit as a computation tool to obtain analytical results. As in the $d=4$ case studied in Ref. \cite{Gubser2009c} we overlook the potential instability induced by the gauge field and proceed to solve analytically eq. \ref{eomAx} in the following three different limits.

%%%%%%%%%%%%%%%%%%%%%%%%%%%
\subsection{Low frequencies}\label{sec:low_freq}
%%%%%%%%%%%%%%%%%%%%%%%%%%%%%%%%%%%%
The small frequency dependence of $\sigma$ is studied by solving eq. (\ref{eomAx}) in the IR limit. The scalar is now locked around its minimum, $\psi_\text{IR}$. By using the asymptotic values of $A$ and $h$ in the IR limit eq.\ref{eomAx} simplifies to,
\begin{equation}\label{eomAxIR}
A_{x}''(r)+         A_{x}'(r)    \frac{d-2}{L_\text{IR}}      +A_{x} (r)   \left(    \frac{\omega^{2}}{e^{2r/L_\text{IR}}}   -  2 q^{2}\psi_\text{IR}^{2}\right)=0, 
\end{equation}
where we have assumed that $e^{\frac{2r}{\LIR}}\phi'(r)\to0$ as $r\to-\infty$. The solution of the above equation can be written in terms of a Hankel function as:
\begin{equation}\label{AxsolIR}
\begin{split}
&A_x(r)=e^{-\frac{d-2}{2\LIR}r}H^{(1)}_{\alpha}\left(\omega\LIR e^{-{r\over\LIR}}\right),\ \ \alpha=\Delta_{\phi_{\text IR}}-{d-2\over 2}=\frac{1}{2}\sqrt{(d-2)^2+8q^2\psi^2_{\text{IR}} \LIR^2}.
\end{split}
\end{equation}
As was pointed out previously, \cite{Gubser2009,Horowitz2009}, the frequency dependence of the conductivity at zero temperature is extracted from the conservation of the flux ($\partial_r \F=0$) with $\F={-he^{(d-2)A}\over 2i}A_x ^*\overleftrightarrow{\partial_r}A_x$,  
\begin{equation}\label{flux}
\Real(\sigma)\propto{ \F \over \omega |A_0|^2}
\end{equation}

Notice that, modulo a factor $i/2$, the flux $\F$ coincides with the definition of $\F(k,z)$ given in eq. (\ref{sgauge}), namely $\F(k,z)=-{\sqrt{-g}g^{zz}g^{xx}\over 2}\tilde A_{x}(z,-k)\partial_z \tilde A_x(z,k),$ where in this case the metric is given by eq. (\ref{metric}). In the latter the holographic coordinate is $r$, instead of $z = 1/r$, and we take the gauge field in position space instead of momentum space.

To obtain $A_0$ we need to match the solution given in eq. (\ref{AxsolIR}) to $Z(r)$, the solution of eq. (\ref{eomAx})  with $\omega=0$, which is assumed to satisfy $Z(r)\to e^{-{r\over \LIR}\left({d-2\over2}-\alpha\right)}$ as $r\to-\infty$. 

For $\omega$ small enough, such that $r^*\ll \rir$, where $r^*=\LIR \log \omega \LIR$ and $\rir$ is the scale at which the geometry is significantly deformed from eq. (\ref{metricIR}), the convergent part of the solution, eq. (\ref{AxsolIR}), is matched to $Z(r)$ in the region $r^*\ll r \ll \rir$:
\begin{equation}
A_x(r)\simeq C\ Z(r)  (\omega \LIR)^{-\alpha},
\end{equation}
where $C$ is a constant. Therefore, taking the limit of the previous expression when $r\to\infty$ results in 
\begin{equation}
A_0\propto \omega^{-\alpha},
\end{equation}
and, from  eq. (\ref{flux}), the conductivity is, 
\begin{equation}\label{eq:small_freq}
\Real(\sigma)\propto \omega^{2\alpha-1},
\end{equation}
with $\alpha=\Delta_{\phi_{\text IR}}-{d-2\over 2}=\frac{1}{2}\sqrt{(d-2)^2+8q^2\psi^2_{\text{IR}} \LIR^2}$. 
The exponent $\alpha$ that controls the strength of the low energy excitations increases with $d$. This is a strong suggestion that, in agreement to the results at finite temperature, high dimensionality suppress low energy excitations and therefore make the system less strongly interacting. The $d$ dependence of the conductivity in low frequency limit was previously investigated in Ref.\cite{zeng2013}. However the expression for the conductivity in \cite{zeng2013} is not the same as eq.(\ref{eq:small_freq}). We note that eq. (\ref{eq:small_freq}) agrees with the results of  Refs.\cite{Gubser2009,Gubser2009c} for $d=3,4$ as well as with our numerical results up to $d=9$. We observe that as $d$ increases, the region where eq.(\ref{eq:small_freq}) is a good approximation is restricted to smaller frequencies. Moreover, since for larger $d$ the divergence in $A_x$ is stronger, see eq. (\ref{AxsolIR}), the numerical results become less reliable, and harder to obtain, in this limit.  

Finally we also note that, in Lifshitz backgrounds with hyperscaling violation, the DC conductivity for small frequencies shows a similar power law behavior \cite{Charmousis2010,Gouteraux2013}. It would be interesting to carry out a $1/d$ expansion in these type of backgrounds in order to explore universal features in the large $d$ limit. 

%%%%%%%%%%%%%%%%%%%%%%%%%%%
\subsection{Large frequencies}\label{sec:large_freq}
%%%%%%%%%%%%%%%%%%%%%%%%%%%%%%%%%%%%
We now explore the large frequency limit of the conductivity corresponding to the region where the frequency $\omega$ is the largest energy scale in the problem, namely, it is much larger than the chemical potential or the condensate. Since the conductivity has units of energy to $d-3$ we expect that in this limit its real part $\propto \omega^{d-3}$. This can be confirmed explicitly by rewriting the prefactor in front of $A_x$ in the third term of the left hand side of eq. (\ref{eomAx}) as:
\begin{equation}\label{Ax_eq_term}
{1\over h}\left({\omega^2\over h}e^{-2A}-{2q^2\psi^2}-{\phi'^2e^{-2A}}\right).
\end{equation}
For $\omega L \to\infty$ and $r\to\infty$ such that $\omega L e^{-r/L}\sim {\cal O}(1)$, the last two terms are negligible with respect to the first, 
\begin{equation}
2q^2\psi(r)^2+e^{-2A(r)}\phi'(r)^2\sim 2q^2\psi_{UV}^2e^{-{r\over L}A_{UV}(d-\Delta_{\psi_{UV}})}+{\rho^2(d-2)^2\over L^2}e^{-2d{r\over L}},
\end{equation}
by virtue of the boundary conditions given in eq. (\ref{bcUV}). These terms are negligible compared to the term $\propto\omega^2$, which by assumption is
\begin{equation}\label{omegaterm_largeomega}
{\omega^2\over h e^{2A}}\sim {\cal O}(1),
\end{equation}
in the region of $r$ considered. For even larger $r$, the previous term becomes arbitrarily small and no additional $\omega$ dependence is introduced. Thus, all the frequency dependence of $A_x$ is obtained, in this region of frequency, by setting the scalar to zero and $h\simeq1$ and $A\simeq r/L$. This leads to
 \begin{equation}\label{Axlargeomega}
A_x(r)=  e^{-\frac{d-2}{2L}r}\left[H^{(1)}_{d-2\over 2}\left(\omega L e^{-{r\over L}}\right)+C_2H^{(2)}_{d-2\over 2}\left(\omega L e^{-{r\over L}}\right)\right],\ \ C_2\in\mathbb{R}.
\end{equation}
$C_2$ has to be determined from the solution in the bulk, however since we set $h\simeq1$ and $A\simeq r/L$ in the whole domain of $r$, the solution in the IR is approximatively given by setting the solution above. Therefore, the ingoing boundary conditions imply $C_2\sim0$. Physically, for large enough $\omega$ the perturbation is insensitive to the flow between the two AdS spaces, and, in particular, to the presence of a nonzero scalar field in the degenerate horizon.

Close to the boundary, eq. (\ref{Axlargeomega}) reads
\begin{equation}\label{Axlargeomegabdy}
A_x(r)\sim C \omega^{2-d \over 2}+\dots,
\end{equation}
where $C$ is a constant and the dots stand for terms which depend on $\omega$ but decay exponentially with $r$. Therefore,
\begin{equation}\label{A0large}
A_0=\lim_{r\to\infty}A_x(r)\propto \omega^{2-d \over2}
\end{equation}
and, as before, using eq. (\ref{flux}) leads to, 
\begin{equation}\label{large_frequency_sigma}
\Real(\sigma)\simeq C_d^{-1} \omega^{d-3}
\end{equation}
with
\begin{equation}
C_d={\pi\over 2}\left[ \Gamma(2-d/2)\right]^{-2} 2^{d-2}.
\end{equation} 
We note that this result is strictly valid for odd dimensions only. The case $d=4$ has been discussed in \cite{Horowitz2008} where, it was found $\sigma\simeq \omega[\pi/2+i(\gamma+\log {\omega\over 2T_c})]$ for large frequencies.

%%%%%%%%%%%%%%%%%%%%%%%%%%%%%%%%%%%%
%%%%%%%%%%%%%%%%%%%%%%%%%%%%%%%%%%%%
\section{Analytical calculation of the entanglement entropy in $\boldsymbol{d \gg 1}$ dimensions}\label{Sec:EE}
%%%%%%%%%%%%%%%%%%%%%%%%%%%%%%%%%%%%
The entanglement entropy is a valuable source of information of strongly interacting systems including the classification of the different quantum phases, the estimation of the effective number of degrees of freedom of the theory, the rate of propagation of information after a perturbation or the location and characterization of phase transitions even in cases where there is no order parameter. In the context of holography it has also been intensively investigated after the landmark conjecture of Ref.\cite{Ryu2006a} provided a relatively straightforward procedure to compute it.  Several papers  \cite{cai2012,cliff2012,pan2014,peng2014,wang2014} have already discussed the entanglement entropy in holographic superconductors \cite{cliff2012,pan2014}, metal-superconductor transitions \cite{peng2014}, metal-insulators transitions \cite{cai2012} or in a superconducting interface \cite{wang2014}. It has been found that the entanglement is a good  observable to characterize these transitions.  Its value is always smaller in the condensed phase and has a discontinuity or a kink (discontinuous derivative) that signals the transition point. It is also sensitive to a mass gap or to the proximity effect in an interface. These calculations in holographic superconductors are numerical as the calculation of the entanglement entropy requires to compute the backreaction of the scalar and gauge fields on the background. The main goal of this section is to show that explicit analytical results are possible in certain $T=0$ backgrounds and also around the critical temperature but only in the limit of large spatial dimensions. This is a strong indication that $1/d$ expansions in holography broadens substantially the scope of the problems that can be addressed analytically. 
  %%%%%%%%%%%%%%%%%%%%%%%%%%%%%%%%%%%%
\subsection{Entanglement entropy at zero temperature}\label{Sec:EET0}
%%%%%%%%%%%%%%%%%%%%%%%%%%%%%%%%%%%%
We now calculate analytically the entanglement entropy at zero temperature related to the background eqs. (\ref{action})-(\ref{hUV}). According to the usual prescription \cite{Ryu2006a} proposed by Takayanagi and Ryu,  given a field theory in $d$ dimensions, the entanglement entropy of a region of space $\tilde A$ and its complement is calculated from the gravity dual by finding the minimal $d-1$-dimensional surface $\gamma_{\tilde A}$ which extends into the bulk such that $\partial \gamma_{\tilde A}=\partial {\tilde A}$. In other words, the boundary of $\gamma_{\tilde A}$ at the AdS$_{d+1}$ boundary is equal to the boundary of $\tilde A$.

To illustrate the calculation we choose ${\tilde A}$ to be a $d-1$ dimensional strip of width $\ell$: ${\tilde A}=\{x\in\mathbb{R}^{d-1}:\ -\ell/2<x^1<\ell/2,\ -a/2<x^i<a/2,\ i=2,\dots,d-2 \}$, where $a$ is the ``length'' of the strip. The entanglement entropy related to the metric eq. (\ref{metric}) is,
\begin{equation}\label{entropy}
S_{\tilde A}=\frac{2a^{d-2}}{4G_N^{d+1}}\int_0^{\ell/2}dx e^{(d-1)A(r)}\sqrt{1+{e^{-2A(r)}\over h(r)}r'(x)^2},
\end{equation}
where, $x=x^1$, $a^{d-2}$ results from integrating $x^i$ with $i=2,\dots,d-2$ and $G_N^{d+1}$ is the $d+1$-dimensional Newton's constant.\\

%%%%%%%%%%%%%%%%%%%%%%%%%%%%%%%%%%%%
\subsubsection{Sharp domain wall approximation}
%%%%%%%%%%%%%%%%%%%%%%%%%%%%%%%%%%%%
As was mentioned above, the background given in eq.(\ref{metric}) interpolates between two copies of AdS space in the IR and UV regions, eqs. (\ref{hIR}) and (\ref{hUV}). Since there is no analytical expression for $h(r)$ and $A(r)$ in the whole range of $r$ we follow \cite{Albash2012} and assume a sharp transition between the two AdS domains at a position denoted by $\rdw$. Numerical results show that there exists a $-\infty<r_m<0$ such that $\psi'(r_m)=0$. It is therefore natural to choose $\rdw=r_m$. Even though we will not be interested in the specific value of $\rdw$ we will require $\rdw<0$ in the following sections. Moreover, numerical results suggest $r_m\propto d^{-1}$.

More specifically the sharp domain wall approximation consists in taking $A(r)$ and $h(r)$ as the asymptotic values given in eq. (\ref{hIR}) for $r<r_{DW}$. Similarly, for $r>r_{DW}$ we take those given in eq. (\ref{hUV}).

As usual in the calculation of $S_{\tilde A}$ with ${\tilde A}$ a strip, eq. (\ref{entropy}) does not depend on the integration variable $x$ explicitly. Therefore, the Euler-Lagrange equations that minimize $S_{\tilde A}$ reduce to the Beltrami identity which states that, given a  Lagrangian $L$,  if $\partial L/\partial x=0$, then $L-r'\partial L/\partial r'$ is a constant. In our case:
\begin{equation}\label{Beltrami_const}
\frac{e^{(d-1) A(r)}}{\sqrt{1 + e^{-2A(r)} r'(x)^2/h(r)}} = \left\{ \begin{array}{lr}
e^{(d-1) \auv  \frac{r_{\ast}}{L}} \ , & r > r_{DW} \\
e^{(d-1) \frac{r_{\ast}}{L_{\rm IR}}} \ , & r < r_{DW} 
\end{array} \right. \ .
\end{equation}
In the previous equation we took into account the different AdS radii, $L$ and $\LIR$ in each region, and $r_*$ is the ``turning'' point of the surface $\gamma_{\tilde A}$ which occurs for $x=0$. We will consider the general case $r_*<r_{DW}$, i.e. the minimal surface extends into the IR region. With the previous considerations eq. (\ref{Beltrami_const}) is easily integrated,
\begin{equation}\label{l2}
\int_0^{\ell/2}dx=\frac{\ell}{2}=I_{\text{IR}}+I_{\text{UV}},
\end{equation}
where
\begin{equation}\label{IIR}
I_{\text{IR}}=\int_{r_*}^{\rdw}{dr\over e^{r\over \LIR} \sqrt{e^{2(d-1){r-r_*\over \LIR}}-1}},\ \ I_{\text{UV}}=\int_{\rdw}^{\ruv}{dr\over \sqrt{\huv} e^{\auv {r\over L}} \sqrt{e^{2(d-1)\auv {r-r_* \over L}}-1}},
\end{equation}
$\ruv$ being the UV cutoff. $I_{\text{IR}}$ is calculated with the change of variables  $t=e^{2(d-2){r-\rdw\over \LIR}}$,
\begin{equation}\label{IIR2}
I_{\text{IR}}=-\LIR e^{-{r_*\over\LIR}}\left[-{\sqrt{\pi}\over d}{\Gamma({3d-2\over2d-2})\over  \Gamma({2d-1\over 2d-2})}+{e^{d{r_*-\rdw\over \LIR}}\over d}\FG{1}{2}{d}{2d-2}{3d-2}{2d-2}{e^{2(d-1){r_*-\rdw\over \LIR}}}\right],
\end{equation}
while an analogous change of variables, $t=e^{2(d-2)\auv{r-\rdw\over L}}$ in $I_{\text{UV}}$ yields
\begin{equation}\label{IUV2}
I_{\text{UV}}= L{e^{[(d-1)r_*-d\rdw]{\auv\over L}}\over d}\FG{1}{2}{d}{2d-2}{3d-2}{2d-2}{e^{2(d-1)\auv{r_*-\rdw\over L}}},
\end{equation}
where we used the relation between $\huv$ and $\auv$ given in eq. (\ref{relation}). Similarly, inserting eq. (\ref{Beltrami_const}) into eq. (\ref{entropy}), the entanglement entropy can be calculated by integrating in $r$ in the two domains ($r<\rdw$ and $r>\rdw$):
\begin{equation}\label{SA}
S_{\tilde{A}}=\frac{2a^{d-2}}{4G_N^{d+1}}\left[S_{\text{IR}}+S_{\text{UV}}\right],
\end{equation}
\begin{equation}
\begin{split}\label{SIR2}	
S_{\text{IR}}=&\int_{r_*}^{\rdw}{e^{(d-2){r\over \LIR}}dr\over\sqrt{1-e^{-2(d-1){r-r_*\over \LIR}}}}={\LIR e^{(d-2){r_*\over \LIR}}\over 2(d-1)}\int_0^1du {y^{1/2}\over u^{1/2}}{1\over{(1-yu)^{3d-4\over 2d-2}}}=\\
&={\LIR e^{(d-2){r_*\over \LIR}}\over 2(d-1)}2\sqrt{y}\FG{1}{2}{3d-4}{2d-2}{3}{2}{y},\\
\end{split}
\end{equation}
where we made the change of variables: $u(r)={1\over y}\left({1-e^{-2(d-1)(r-r_*)/\LIR}}\right)$, $y=1-e^{-2(d-1)(\rdw-r_*)/\LIR}$. eq. (\ref{SIR2}) can be rewritten using the following Hypergeometric function identities,
\begin{equation}\label{Rel1}
\begin{split}
\FF(a,b,c,z)=&{\Gamma(1-a)\Gamma(1-b)(1-z)^{c-a-b}\over\Gamma(1-c)\Gamma(c-a-b+1)}\FF(c-a,c-b,c-a-b+1,1-z)+\\
&+{\Gamma(1-a)\Gamma(1-b)\Gamma(c)\over\Gamma(2-c)\Gamma(c-a)\Gamma(c-b)}z^{1-c}\FF(a-c+1,b-c+1,2-c,z),
\end{split}
\end{equation}
with $b=0,\ c={3d-4\over2d-2},\ a=c-1/2$ and
\begin{equation}\label{Rel2}
c\FF(a-1,b,c,z)-c\FF(a.b-1,c,z)-(a-b)\FF(a,b,c+1,z)=0,
\end{equation}
with $a={1/2},\ b=c={d\over2d-2}$, as follows
\begin{equation}\label{SIR}
\begin{split}
S_{\text{IR}}=&\LIR \left[-{\sqrt{\pi}e^{(d-2){r_*\over \LIR}}\over d(d-2)}{\Gamma({3d-2\over2d-2})\over  \Gamma({2d-1\over 2d-2})}+{e^{(d-2){\rdw\over\LIR}}\over d-2}\sqrt{1-e^{-2(d-1){\rdw-r_*\over\LIR}}}+\right.\\
&\left.+{e^{-d{\rdw\over\LIR}+2(d-1){r_*\over \LIR}}\over d(d-2)}\FG{1}{2}{d}{2d-2}{3d-2}{2d-2}{e^{2(d-1){r_*-\rdw\over \LIR}}}\right],\\
\end{split}
\end{equation}
On the other hand, for $S_{\text{UV}}$, one must take care of the usual divergence for $r\to\infty$. Defining the auxiliary variables $t=e^{-2(d-1)\auv{r-\rdw\over L}}$, the cutoff in the $t$ variable $\tuv=e^{-2(d-1)\auv{\ruv-\rdw\over L}}$ and $y=e^{-2(d-1)\auv{\rdw-r_*\over L}}$ we integrate $S_{\text{UV}}$:\\
\begin{equation}\label{SUV}
\begin{split}
S_{\text{UV}}&={1\over\sqrt{\huv}}\int_{\rdw}^{\ruv}{e^{(d-2)\auv{r\over L}}dr\over\sqrt{1-e^{-2\auv(d-1){r-r_*\over L}}}}={L e^{(d-2)\auv{\rdw\over L}}\over 2(d-1)}\int_{t_{\text{UV}}}^1 {dt\over t^{3d-4\over 2d-2}}{1\over\sqrt{(1-yt)}}=\\
&={L e^{(d-2)\auv{\rdw\over L}}\over 2(d-1)}\left[-2y^{d-2\over2(d-1)}\sqrt{1-ty}\FG{1}{2}{3d-4}{2d-2}{3}{2}{1-yt}\right]^{t=1}_{t=t_{\text{UV}}}=\\   
&=L \left[{e^{(d-2)\auv{\ruv\over L}}\over d-2}-{e^{(d-2)\auv{\rdw\over L}}\over d-2}\sqrt{1-e^{-2(d-1)\auv{\rdw-r_*\over L}}}+\right.\\
&\left.-{e^{[-d\rdw+2(d-1)r_*]{\auv\over L}}\over d(d-2)}\FG{1}{2}{d}{2d-2}{3d-2}{2d-2}{e^{2(d-1)\auv{r_*-\rdw\over L}}}\right].
\end{split}
\end{equation}
In the last equality we used the relations given in eqs. (\ref{Rel1}) and (\ref{Rel2}) and left the cutoff $\ruv$ explicit in the divergent term.

We stress eqs. (\ref{SA}), (\ref{SIR}) and (\ref{SUV}) are an approximation to the entanglement entropy between a strip of width $\ell$ and its complement in the $d$-dimensional boundary when the scalar field condensates. It is interesting to compare these results with the entanglement entropy between a strip of the same width $\ell$ and its complement in the situation in which the scalar is absent, \cite{Ryu2006}. To do so we should express $S_{\tilde{A}}$ in terms of $\ell$, however, from eqs. (\ref{l2}), (\ref{IIR2}) and (\ref{IUV2}) it is clear that $r_*$ cannot be expressed in terms of $\ell$ in a closed form and thus the comparison cannot be made easily. Instead, in the next section we make this comparison only in UV and IR limits of $S_{\tilde{A}}$. Additionally, we also study the large-$d$ limit of $S_{\tilde{A}}$.

%%%%%%%%%%%%%%%%%%%%%%%%%%%%%%%%%%%%
\subsubsection{UV, IR and large-d limits}
{\bf UV limit:} we first consider $r_*>\rdw$, i.e. the minimal surface $\gamma_{\tilde A}$ is embedded in the AdS copy that contains the boundary $r\to\infty$. In this situation $I_{\text{IR}}=S_{\text{IR}}=0$ and $\rdw=r_*$ in eqs. (\ref{IUV2}) and (\ref{SUV}):
\begin{equation}\label{UVlimit}
\begin{split}
&\hspace{4cm}{\ell\over2}=L e^{-\auv{r_*\over L}} {\sqrt{\pi}\over d}{\Gamma({3d-2\over2d-2})\over  \Gamma({2d-1\over 2d-2})},\\ 
&S_{\tilde{A}}^{\text{UV}}\sim\frac{2a^{d-2}L}{4G_N^{d+1}}\left\{{e^{(d-2)\auv{\ruv\over L}}\over d-2}-\left({2\over \ell}\right)^{d-2}{{L^{d-2}\pi^{d-1\over2}}\over d -2}\left[{\Gamma\left({d\over2d-2}\right)\over \Gamma\left({1\over 2d-2}\right)}\right]^{d-1}\right\}.
\end{split}
\end{equation}

As was expected we recover the result for the infinite strip in an AdS space, found in \cite{Nishioka2009}. It is observed the strip width tends to zero following $e^{-\auv{r_*\over L}}$, while the ``finite'' part of the entanglement entropy diverges as $e^{(d-2)\auv{r_*\over L}}$.\\

{\bf IR limit:} In case $r_*\ll\rdw$, i.e. the $\gamma_{\tilde A}$ extends deeply into the IR region. From eqs. (\ref{IIR2}), (\ref{IUV2}),  (\ref{SIR}) and (\ref{SUV}) 
\begin{equation}\label{IRlimit}
\begin{split}
&\hspace{4cm}{\ell\over2}=\LIR e^{-{r_*\over \LIR}} {\sqrt{\pi}\over d}{\Gamma({3d-2\over2d-2})\over  \Gamma({2d-1\over 2d-2})},\\ &S_{\tilde{A}}^{\text{IR}}\sim\frac{2a^{d-2}}{4G_N^{d+1}}\left\{L{e^{(d-2)\auv{\ruv\over L}}\over d-2}-\left({2\over \ell}\right)^{d-2}{{\LIR^{d-1}\pi^{d-1\over2}}\over d -2}\left[{\Gamma\left({d\over2d-2}\right)\over \Gamma\left({1\over 2d-2}\right)}\right]^{d-1}+\right.\\
&\hspace{1cm}\left.+{\LIR e^{(d-2){\rdw\over\LIR}}-Le^{(d-2)\auv{\rdw\over L}}\over d-2}\right\}.\\
\end{split}
\end{equation}
In this limit, the strip width, $\ell$, diverges and the finite part of the entanglement entropy saturates to a constant value given by the first term in the following expression:
\begin{equation}\label{IRlimit2}
\lim_{r_*\to-\infty}S_{\tilde{A}}^{\text{IR}}=\frac{2a^{d-2}}{4G_N^{d+1}(d-2)}\left[\LIR e^{(d-2){\rdw\over\LIR}}-L e^{(d-2)\auv{\rdw\over L}}\right]+\frac{2a^{d-2}}{4G_N^{d+1}}{e^{(d-2)\auv{\ruv\over L}}\over d-2}.
\end{equation}

At this point, it is easy to compare eq. (\ref{IRlimit}) to the entanglement entropy between the strip $\tilde A$ (of same width $\ell$) and its complement in the d-dimensional AdS boundary when $\psi=0$. Were the scalar field be zero, there would be a single AdS space and $S_{\tilde{A}}$ would be given by the first two terms of eq. (\ref{IRlimit}) while the last term would be zero for all $d$. In the presence of the condensate, the third term in the previous equation corresponds to the contribution of the  flow from one AdS copy to the other. Indeed it is easy to see that this term is negative. From the definition of $\LIR$: $-d(d-1)/\LIR^2\equiv V(|\psi_{\text{IR}}|)=-d(d-1)/L^2-m^4/(2u)$, it follows $\LIR<L$. Since we require $\rdw<0$, the term in square brackets of eq. (\ref{IRlimit2})
\begin{equation}
\LIR e^{(d-2){\rdw\over\LIR}}-Le^{(d-2)\auv{\rdw\over L}}< e^{(d-2)\auv{\rdw\over L}}(\LIR -L)<0.
\end{equation}

The conclusion is that the entanglement entropy between a strip of length $\ell$ and its complement is lower if the scalar is present. This means the theory has less degrees of freedom in this case. In the limit of a strip of infinite width ($\ell\to\infty$), the finite contribution of $S_{\tilde{A}}$ reaches the maximum value given by the first term in eq. (\ref{IRlimit2}). These results are expected as the entanglement entropy counts the degree of freedom of the theory. It is therefore natural that it is smaller in the condensed phase.

Let us turn to the study of the large-d limit of $S_{\tilde A}$. Before we do so, we must analyze the behavior of the strip length as $d\to\infty$. From eqs. (\ref{l2}), (\ref{IIR2}) and (\ref{IUV2}) it is clear that if $r_*$ either remains constant or increases, as $d$ increases, both $I_{\text{IR}}$ and $I_{\text{UV}}$ would tend to zero and $l\to0$. In order to compare $S_{\tilde{A}}$ for different $d$ we must keep $\ell$ constant. Therefore, $r_*\to-\infty$, as $d\to\infty$, which corresponds to the IR limit ($r_*\ll \rdw$) studied above. Taking $d$ large and $\ell$ constant in eq. (\ref{IRlimit}) yields,
\begin{equation}\label{r_l}
r_*(d\to\infty)\sim-\LIR \log{\ell d\over\pi\LIR},
\end{equation}
and 
\begin{equation}\label{Slarged_final}
\begin{split}
S_{\tilde{A}}(d\to\infty)\sim\frac{1}{4G_N^{d+1}}&\left[{2a^{d-2}Le^{(d-2)\auv{\ruv\over L}}\over d-2}-{\pi^{d-1}\LIR^{d-1}\over (d-2)(d-1)^{d-1}}\left({a\over \ell}\right)^{d-2}+\right.\\
&\left.+2a^{d-2}{\LIR e^{(d-2){\rdw\over\LIR}}-Le^{(d-2)\auv{\rdw\over L}}\over d-2}\right].\\
\end{split}
\end{equation}
The second term of the previous equation corresponds to the universal contribution for the infinite strip in an AdS space, \cite{Nishioka2009} which is strongly suppressed in the $d \to \infty$ limit as it is proportional to $d^{-d}$. The third term has some interesting features. It is independent of $\ell$ as it is expected in gapped systems where the typical length, the numerator in this case, is closely related to the coherence length of the holographic superconductor. Its $d$-dependence, arising from the flow of one AdS space to another, is dictated by the $d$-dependence of $\rdw$ and $\auv$. As mentioned before, numerical results for $d\leq9$ suggest  ${\rdw\over \LIR}\propto d^{-1}$ and $\rdw<0$ which implies a behavior like $d^{-1}$ for these contributions. In the limit of a vanishing scalar field, the third term vanishes for all $d$ and the result of Ref.\cite{Nishioka2009} is recovered. 

Let us simplify eq. (\ref{Slarged_final}) for the particular set of parameters: $m^2L^2=-3d^2/16$, $\psi_{\text{IR}}=\sqrt{d-1\over d}$, $u=-{m^2L^2\over \psi^2_{\text{IR}}}$, $L=1$. These values, together with the definition of $L_{\text{IR}}$: ${-d(d-1)\over \LIR^2}\equiv V(|\psi_{\text{IR}}|)$, yield a constant, in $d$, $L_{\text{IR}}=\sqrt{32/35}L$. Moreover, as discussed earlier, ${A_{\text{UV}}\over L}<{1\over L_{\text{IR}}}$. These considerations allow a further simplification of eq. (\ref{Slarged_final}),
\begin{equation}\label{Slarged_final1}
\begin{split}
S_{\tilde{A}}&(\ell,a,d\to\infty)\lesssim\frac{a^{d-2}}{4G_N^{d+1}}\left[S_{div}-{\pi^{d}\LIR^{d-1}\over d^{d}}{1\over \ell^{d-2}}+2e^{{\alpha\over \LIR}}{\LIR -L\over d}\right]\\
&\sim\frac{a^{d-2}}{4G_N^{d+1}}\left[S_{div}-{\pi^{d}\over d^{d}}\left({32\over35}\right)^{d-1\over2}{1\over \ell^{d-2}}+2\Delta Le^{{\alpha}}{1\over d}\right],\\
\end{split}
\end{equation}
where $\ell$ and $a$ are the width and the characteristic length (infinite) of the transverse dimensions of the strip. The radius of curvature of the IR asymptotic AdS space, $\LIR$, does not depend on $d$, $\alpha$ is the constant of proportionality in $\rdw\simeq{\alpha\over\LIR d}$ which can only be obtained numerically and $\Delta L=\LIR-L=\sqrt{32/35}-1<0$. $S_{div}$ is the (divergent) part containing the UV integration cutoff.

As we mentioned previously, were the condensate vanish, the last term in eq. (\ref{Slarged_final1}) would be identically zero, since the asymptotic IR and UV AdS radii would be the same, $\LIR=L$. Moreover, this term is negative, which means the finite part of the entanglement entropy is lower, and thus indicates less degrees of freedom in the presence of the condensate. Finally, as $d\to\infty$, this contribution is smaller, suggesting the difference between the entanglement entropy in the presence and absence of the condensate is smaller. The latter is an indication that, in agreement with the conductivity results, the condensate  interactions become weaker as $d$ increases.
%%%%%%%%%%%%%%%%%%%%%%%%%%%%%%%%%%%%
\subsection{Entanglement entropy close to the transition}\label{sec:EE_Tc}
%%%%%%%%%%%%%%%%%%%%%%%%%%%%%%%%%%%%
In this section we compute analytically the entanglement between the semi-infinite strip, $\tilde A$, defined in the previous section and its complement at finite temperature. We employ the action eq.(\ref{action2}) but we have to go beyond the probe limit. We assume the following parametrization of the metric:
\begin{equation}
ds^{2}={1\over L^2 z^2}\left(-f(z)e^{-\chi(z)} dt^2+{L^4\over f(z)}dz^2+dx_i^2\right),
\end{equation}
where $i=1,\dots,d-1$, $z=1/r$. Above the transition, the metric corresponds to the AdS planar Reissner-Nordstr\"{o}m in $d+1$ dimensions. More precisely, $\chi(z)=0$, the gauge field $A_t=\phi=\mu[1-(z/z_0)^{d-2}]$ and $f(z)=\frn\equiv1-(1+Q^2) {\left(z\over z_0\right)}^d+Q^2{\left(z\over z_0\right)}^{2d-2}$, where $Q^2={\mu^2 z_0^2\gamma^{2}}$, $\gamma^2={d-2\over d-1}{L^4\over 2}$ and $z_0$ is the inverse of the outer horizon $r_0$.

Throughout this section we take $d$ to be large so we can get explicit analytical results. We also consider strips of length $\ell$ for which the minimal surface $\gamma_{\tilde A}$, associated to the strip, does not extend too deeply into the bulk, such that the turning point, $z_*$, satisfies $(z_*/z_0)^d \ll  1$. This is in general a good approximation in the $d \to \infty$ limit, even for $z_*\lesssim z_0$.
Moreover we restrict ourselves to the region  $T\sim T_c$ and therefore, the dual order parameter $\langle {\cal O}\rangle$ is very small compared to the typical energy scale $T_c$. This regime restricts the generality of the results for the entanglement entropy but allows to estimate analytically the correction in the presence of the scalar field close to the phase transition.

The entanglement between the strip and its complement is given by:
\begin{equation}\label{SAT}
\begin{split}
s_{\tilde A}&\equiv S_{\tilde A}\frac{4G_N^{d+1}}{a^{d-2}L^{d-1}}={2}\int_0^{\ell/2}dx {1\over z^{d-1}}\sqrt{1+{z'(x)^2\over f(z)}}=\\
&=2z_*^{d-1}\int_{\zuv}^{z_*}{dz\over z^{d-1}}{1\over\sqrt{f(z)\left(z_*^{2d-2}-z^{2d-2}\right)}},\\
\end{split}
\end{equation}
where we have rescaled $z\to z/L^2$ in order to compare with the results in Ref.\cite{cliff2012}. We have also introduced the UV cutoff $\zuv\to0$ and, as before, we have used the fact that the integral does not depend on $x$.  The turning point, $z_*$, of the surface $\gamma_{\tilde A}$ embedded into the bulk is given by $ z_*^{d-1}= z^{d-1}\sqrt{1+z'(x)^2/f(z)}$. The strip width, $\ell$ is related to $z_*$ as follows:
\begin{equation}\label{lT}
\frac{\ell}{2}=\int_0^{z_*}dz {z^{d-1}\over\sqrt{f(z)\left(z_*^{2d-2}-z^{2d-2}\right)}}.
\end{equation}

Even in the absence of the scalar field in eq.(\ref{action}), i.e., the Reissner-Nordstr\"{o}m background, the previous two integrals cannot be computed analytically for arbitrary $d$. However, an analytical calculation is possible in the large $d$ limit.

First, we calculate the width of the strip from eq. (\ref{lT}), by setting $f(z)=\frn(z)$ and expanding $\sqrt{\frn(z)}$ in powers of $z/z_0$:\footnote{For simplicity it is more convenient to  expand in $\delta=-(1+Q^2) {\left(z\over z_0\right)}^d+Q^2{\left(z\over z_0\right)}^{2d-2}$.}
\begin{equation}\label{lseries}
\frac{\ell}{2}=  {z_*\sqrt{\pi}\over d}{\Gamma\left({3d-2\over 2d-2}\right)\over \Gamma\left({2d-1\over 2d-2}\right)}+z_*\sum_{n=1}^{\infty}\sum_{l=0}^{n}C_{nl} \alpha^l(-\beta)^{n-l}\left(z_*\over z_0\right)^{b_{nl}},
\end{equation}
where,
\begin{equation}\label{Cnl}
C_{nl}={(2n-1)!!\over 2^n(n-l)! l!}{\Gamma\left({2d+a_{nl}-1\over 2d-1}\right)\over \Gamma\left({a_{nl}+d\over 2d-2}\right)}{\sqrt{\pi}\over a_{nl}+1},
\end{equation}
$\alpha=1+Q^2$, $\beta=Q^2$ and $a_{nl}=2dn+d(1-l)+2(l-n)-1$,  $b_{nl}=(2d-2)(n-l)+dl$.  In the large $d$ limit, assuming $\ell$ fixed, it is enough to keep only the terms corresponding to $n=1$ in the series above. The resulting expression of the strip length, $\ell$, as a function of the turning point of the minimal surface, $z_*$, is,
\begin{equation}\label{lRN}
\frac{\ell}{2}\simeq {z_*\over d}\left[{\pi\over 2}+{1+Q^2\over 2d}\left({z_*\over z_0}\right)^d-{Q^2\pi\over 8d}\left({z_*\over z_0}\right)^{2d-2}+\dots\right]. 
\end{equation}
Similarly, from eq. (\ref{SAT}), with  $f(z)=\frn(z)$,
\begin{equation}\label{sseries}
s_{\tilde A}={2\over (d-2)\zuv^{d-2}}-{2\sqrt{\pi}\over(d-2)z_*^{d-2}}{\Gamma\left({d\over 2d-2}\right)\over\Gamma\left({1\over 2d-2}\right)}+{2\over z_*^{d-2}}\sum_{n=1}^{\infty}\sum_{l=0}^{n}C_{nl} \alpha^l(-\beta)^{n-l}\left(z_*\over z_0\right)^{b_{nl}},
\end{equation}
where $C_{nl}$ is given in eq. (\ref{Cnl}), $a_{nl}=(2d-2)(n-l)+d(l-1)+1$, $b_{nl}=(2d-2)(n-l)+dl$. For large $d$, taking the first two terms of the series,
\begin{equation}\label{sRN}
\begin{split}
s_{\tilde A}&\simeq\frac{2}{d\zuv^{d-2}}-\frac{\pi}{d^2z_*^{d-2}}+{1+Q^2\over 2z_*^{d-2}}\left({z_*\over z_0}\right)^d-{\pi Q^2\over2z_*^{d-2}}\left({z_*\over z_0}\right)^{2d-2}+\dots\\
&\simeq\frac{2}{d\zuv^{d-2}}-\frac{\pi^{d-1}}{d^d}{1\over \ell^{d-2}}+{1+Q^2\over 2 z_0^{d}}{d^2\over\pi^2} \ell^2-{ Q^2\over2\pi^{d-1} z_0^{2d-2}}{d^d  }\ell^d+\dots,\\
\end{split}
\end{equation}
 where, in the last equality we substituted $z_*\sim{d \ell\over \pi}$, which is a good approximation as long as $z_*\ll z_0$ (small $\ell$) and $d$ fixed or, for a fixed $z_*\lesssim z_0$, and sufficiently large $d$. In the latter case, $\ell$ should also be small, which means that as $d$ increases the minimal surface $\tilde A$ should reach the near-horizon region for smaller strip lengths. $Q$ is related to the chemical potential and the position of the outer horizon, through $\mu$ and $z_0$, $Q^2={\mu^2 z_0^2\gamma^{2}}$, $\gamma^2={d-2\over d-1}{L^4\over 2}$.
 
In the presence of the scalar field, $\psi$, analytical results are harder to obtain close to the phase transition $T \lesssim T_c$ since $f(z)$ is subject to the backreaction of $\psi$, and, in general, cannot be written in a closed form. 

However, we show below that it is still possible to find an explicit analytical expression in the large-$d$ limit. 

In order to proceed we  solve perturbatively the equations of motion close to the transition. To do so we expand the fields in the equations of motion (see the appendix \ref{App_EE}  for more details) in a power series in a quantity related to the VEV of the operator dual to the scalar field. More specifically, from the UV boundary condition of the scalar field, $\psi\sim {\alpha\over r^{\Delta_{-}}}+{\beta\over r^{\Delta_{+}}}+\dots$, given in eq. (\ref{bdy_cond}), we set $\alpha=0$ and {\it define} $\ep\equiv\beta$. Close to the transition this expansion parameter is related to temperature in the usual way, $\epsilon^{\Delta_+}\propto\langle \OO\rangle\propto (T-T_c)^{1/2}$, with $\Delta_+$ being the conformal dimension. \\
The blackening function can be expanded as $f(z)\simeq \frn+\ep^2(f_2^a(z)+\dots)$ with $f_2^a(z)=-{\mu_0\kappa  z_0^2}\left[\left({z\over z_0}\right)^{d}-\left({z\over z_0}\right)^{2d-2}\right]$ and the dots indicate subleading terms, see  appendix eq. (\ref{app:f}), where $\mu_0$ is the chemical potential at the phase transition and $\kappa$ is an integration constant which is calculated from the perturbative analysis of the equations of motion, eq. (\ref{app:deltamu}). It is negative $\kappa <0$ for $d\ge3$.

The calculation of the entanglement entropy including the leading correction $\ep^2f_2^a(z)$ is totally analogous to the one corresponding to the Reisnner-Nordstr\"{o}m case given in detail above. The main difference is that $\alpha$ and $\beta$ in eqs. (\ref{lseries}) and (\ref{sseries}) are replaced by,
\begin{equation}\label{alphabeta_conds}
\begin{split}
&\tilde \alpha=1+\tilde Q^2+\epsilon^2\mu_0\kappa  \tilde z_0^2,\ \ \tilde \beta=\tilde Q^2+\epsilon^2\mu_0\kappa \tilde z_0^2.\\
\end{split}
\end{equation}
Here, $\tilde Q^2=\mu_0^2\tilde z_0^2\gamma^2\ne Q^2$ and $\tilde z_0\ne z_0$, in order to take into account the different horizon radius with respect to a pure Reissner-Nordstr\"{o}m black hole at the same temperature.

Consequently, in the large-$d$ limit, the relation between the strip width and the turning point of $\gamma_{\tilde A}$ in the hairy black hole background is,
\begin{equation}\label{lback}
\frac{\ell}{2}\simeq{\tilde z_*\over d}\left[{\pi\over 2}+{\tilde \alpha\over 2d}\left({\tilde z_*\over \tilde z_0}\right)^d-{\tilde \beta \pi\over 8d}\left({\tilde z_*\over \tilde z_0}\right)^{2d-2}+\dots\right] 
\end{equation}
Similarly, $\tilde s_A$ is
\begin{equation}\label{sback}
\begin{split}
\tilde s_{A}\simeq &\frac{2}{d\zuv^{d-2}}-\frac{\pi}{d^2\tilde z_*^{d-2}}+{1+\tilde Q^2+\mu_0\kappa  \epsilon^2\tilde z_0^2\over 2\tilde z_*^{d-2}}\left({\tilde z_*\over \tilde z_0}\right)^d-\pi{\tilde Q^2+\mu_0\kappa  \epsilon^2 \tilde z_0^2\over2\tilde z_*^{d-2}}\left({\tilde z_*\over \tilde z_0}\right)^{2d-2}+\dots,
\end{split}
\end{equation}
where, $\kappa<0$ is given in the appendix \ref{App_EE}. In order to compare the entanglement entropy between the strip and its complement in the condensed phase with the one in the symmetry unbroken phase one needs, in principle, to compute the charge, $Q$, and horizon position, $z_0$, of a Reisnner-Nordstr\"{o}m black hole at the same temperature, eq. (\ref{eq:Qz0RN}). However it is important to note that the contribution due to the condensate, contained in the $\epsilon^2$ term, always leads to less entanglement in the condensed phase ($\mu_0>0$ and $\kappa<0$). 

To compute the Reissner-Nordstr\"{o}m black hole parameters at the same temperature as the hairy black hole we fix the horizon in the superconducting phase, $\tilde z_0=1$, and solve the following equations in the horizon, $z_0$, and charge, $Q$:
\begin{equation}\label{eq:Qz0RN}
\mu={Q\over z_0\gamma},\ \ {T\over \rho^{1\over d-1}}={1\over 4\pi}{d-(d-2)Q^2\over \left(Q/\gamma\right)^{1\over d-1}},
\end{equation}
where $T/\rho^{1\over d-1} $ is a function of $\epsilon$  (proportional to $\langle \OO\rangle^{1/\Delta}$), the metric components at the horizon and the chemical potential at the the transition, $\mu_0$, eq. (\ref{app:T}), and $\mu=\mu_0+\epsilon^2(\kappa +\phi_2^b(0))>\mu_0$, where $\kappa $ and $\phi_2^b(0)$ are integration constants given in the App. \ref{App_EE}. From eq. (\ref{eq:Qz0RN}) it follows that $\tilde z_0>z_0$ and $\tilde Q<Q$, and solving eqs. (\ref{lRN}) and (\ref{lback}) one obtains $\tilde z_*>z_*$. Consequently, comparing the finite contributions in eqs. (\ref{sRN}) and (\ref{sback}), we conclude that below, but close, to the phase transition the number of degrees of freedom in the dual field theory is smaller than in the normal phase (no condensate, $\epsilon=0$, $\mu=\mu_0$). This is again consistent with the theoretical expectation that the entanglement entropy is closely related to the effective number of degrees of freedom of the system at a given temperature.  

For completeness, we express $\tilde s_{\tilde A}$ in terms of the strip length $\ell$, the expansion parameter $\epsilon\propto\langle\OO\rangle^{1/\Delta}$, $\mu_0$ and $\kappa$. From eq. (\ref{lback}), $\tilde z_*\sim {d \ell\over \pi}\left[1+{\cal O}(d^{-2d})\right]$. Substituting $z_* = {d \ell\over \pi }$ in eq. (\ref{sback}), the final expression of the entanglement entropy in terms of the strip length is given by,
\begin{equation}\label{ee_hairy_l}
\begin{split}
\tilde s_{\tilde A}\simeq &\frac{2}{d\zuv^{d-2}}-\frac{\pi^{d-1}}{d^d}{1\over \ell^{d-2}}+{1+\tilde Q^2+\mu_0\kappa  \ \epsilon^2\tilde z_0^2 \over 2\tilde z_0^{d}}{d^2\over\pi^2} \ell^2-{ \tilde Q^2+\mu_0\kappa \ \epsilon^2\tilde z_0^2\over2\pi^{d-1}\tilde z_0^{2d-2}}{d^d}\ell^d+\dots.\\
\end{split}
\end{equation}

Before we compare eq.(\ref{ee_hairy_l}) with numerical results we discuss the limits of applicability of the linear approximation  $z_* \sim  \tilde z_* \propto \ell$.

\begin{figure}[H]
\center
\includegraphics[scale=0.7]{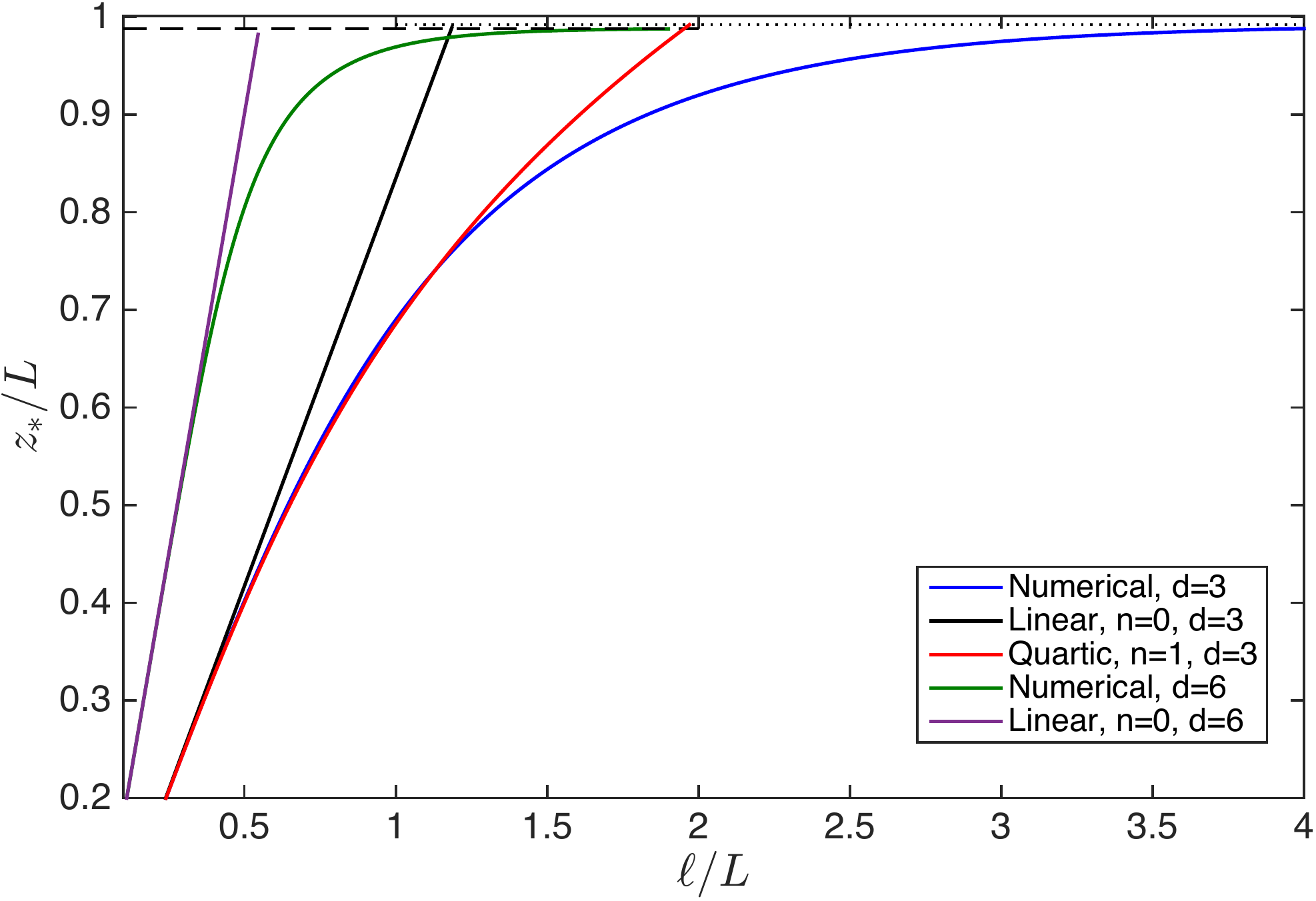}
\vspace{-0.5cm}
\caption{Position of the tip, $z_*$, of the minimal surface in the Reissner-Nordstr\"{o}m  background for $d=3$ and $d=6$. For $d=3$, $\mu=2.02$, $Q^2=\mu^2z_0^2\gamma^2$, $\gamma^2={d-2\over d-1}{L^4\over 2}$ and $z_0=0.992$ (dotted line), while for $d=6$, $\mu=0.38$ and $z_0=0.988$ (dashed line). The ''numerical'' results are obtained from the numerical integration of eq. (\ref{lT}) with $f(z)=\frn$. The linear approximation $z_*\propto \ell$ corresponds to $n=0$ in eq. (\ref{lseries}) and $\alpha=1+Q^2$, $\beta=Q^2$. The analytical solution of the fourth degree polynomial in $z_*$ contains the leading correction, the first term of the series given in eq. (\ref{lseries}). The linear approximation $z_*\propto \ell$ is clearly only valid for small $\ell$ and, for larger $d$, it becomes gradually better deep in the bulk. Including more corrections in higher powers of $z_*/z_0$ gives a better approximation but requires, in general, numerical methods. }
\label{fig:lRN3d}
\end{figure}

\begin{figure}[H]
\center
\includegraphics[scale=0.7]{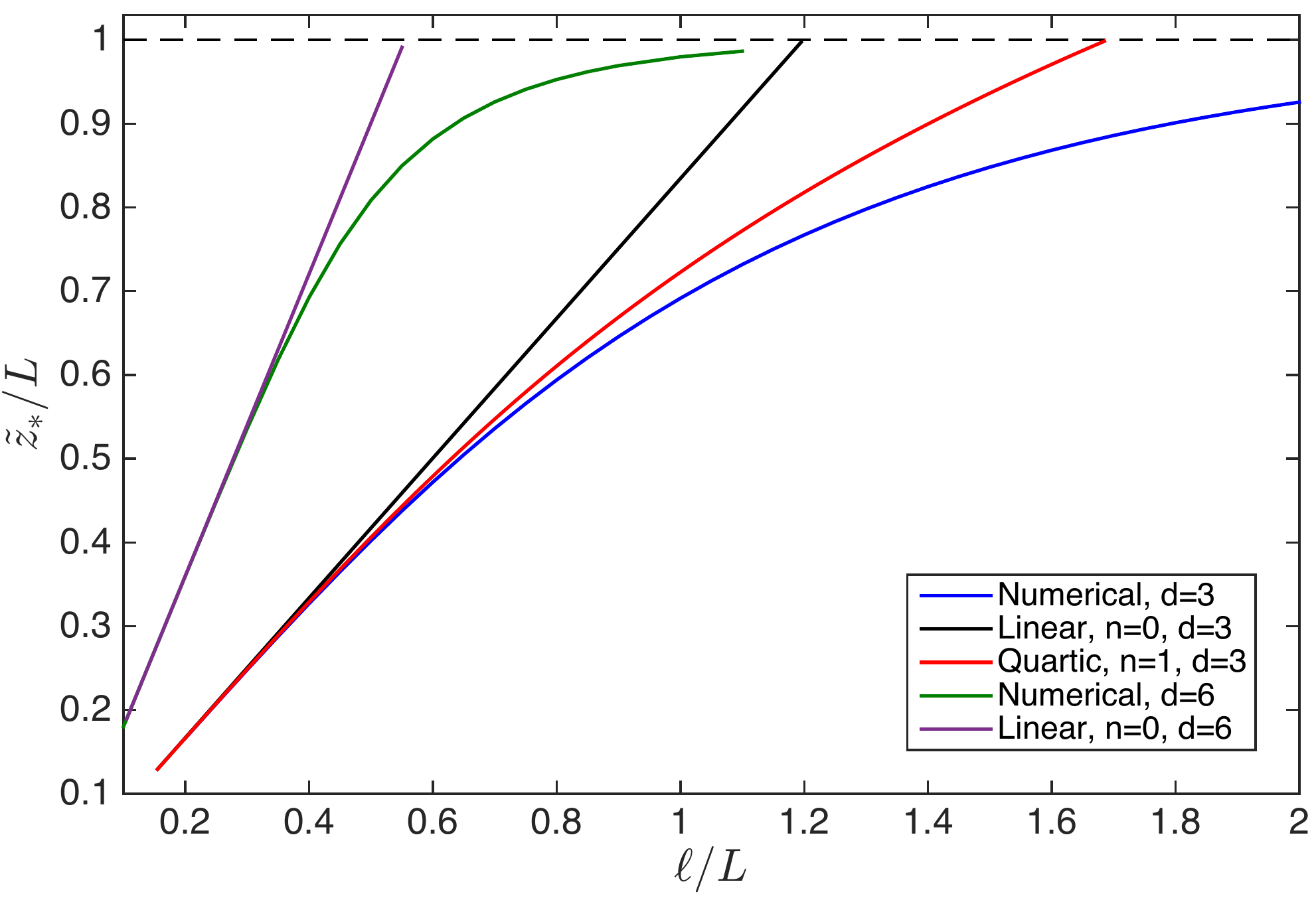}
\vspace{-0.5cm}
\caption{Position of the tip, $\tilde z_*$,  of the minimal surface in the superconducting phase for $d=3$, $\mu=2.00$ and $d=6$, $\mu=0.37$. In both cases the horizon is fixed at $z_0=1$ (dashed line) and $Q^2=\mu_0^2{\tilde z_0^2}\gamma^2$, $\gamma^2={d-2\over d-1}{L^4\over 2}$. Similarly to figure  \ref{fig:lRN3d} the numerical results are obtained from eq. (\ref{lT}) with $f(z)=\frn+\epsilon^2 f_2$ and $f_2$ given in eq. (\ref{app:f}). The rest of the lines are obtained by truncating the series in eq. (\ref{lseries}) at linear and quartic powers of $z_*$ with $\alpha$ and $\beta$ given in eq. (\ref{alphabeta_conds}). Similarly to the symmetry unbroken phase, figure  \ref{fig:lRN3d}, the linear approximation $\tilde z_*\propto \ell$ is better for larger $d$.}
\label{fig:lH3d}
\end{figure}

In figures \ref{fig:lRN3d},\ref{fig:lH3d} we depict the dependence of the tip of the surface $\tilde A$ on $\ell$ resulting from solving eqs.(\ref{lT}) and (\ref{lseries}).
For small $\ell$, the linear approximation agrees well with the exact result,  eq. (\ref{lT}). 
As $\ell$ grows this agreement worsens substantially. Additionally, as $d$ increases, the approximation $z_*=\tilde z_* =\ell {d\ \Gamma\left({2d-1\over 2d-2}\right)\over 2\sqrt{\pi} \Gamma\left({3d-2\over 2d-2}\right)}$ is valid for smaller values of $\ell$ but, at the same time, since the corrections $\OO(d^{-2d})$ are smaller, it remains a good approximation closer and closer to the horizon for both the normal ($T<T_c$), figure  \ref{fig:lRN3d}, and the condensed phase close to the transition, figure  \ref{fig:lH3d}. This is nothing else but a consequence of the simplification of general relativity in the large-$d$ limit. For a black hole, as dimensionality increases, its region of influence shrinks to a neighbourhood of the horizon \cite{Emparan2013}. 
\vspace{0mm}
\vspace{-1cm}

\begin{figure}[H]
\hspace{-7mm}
\includegraphics[scale=1.2]{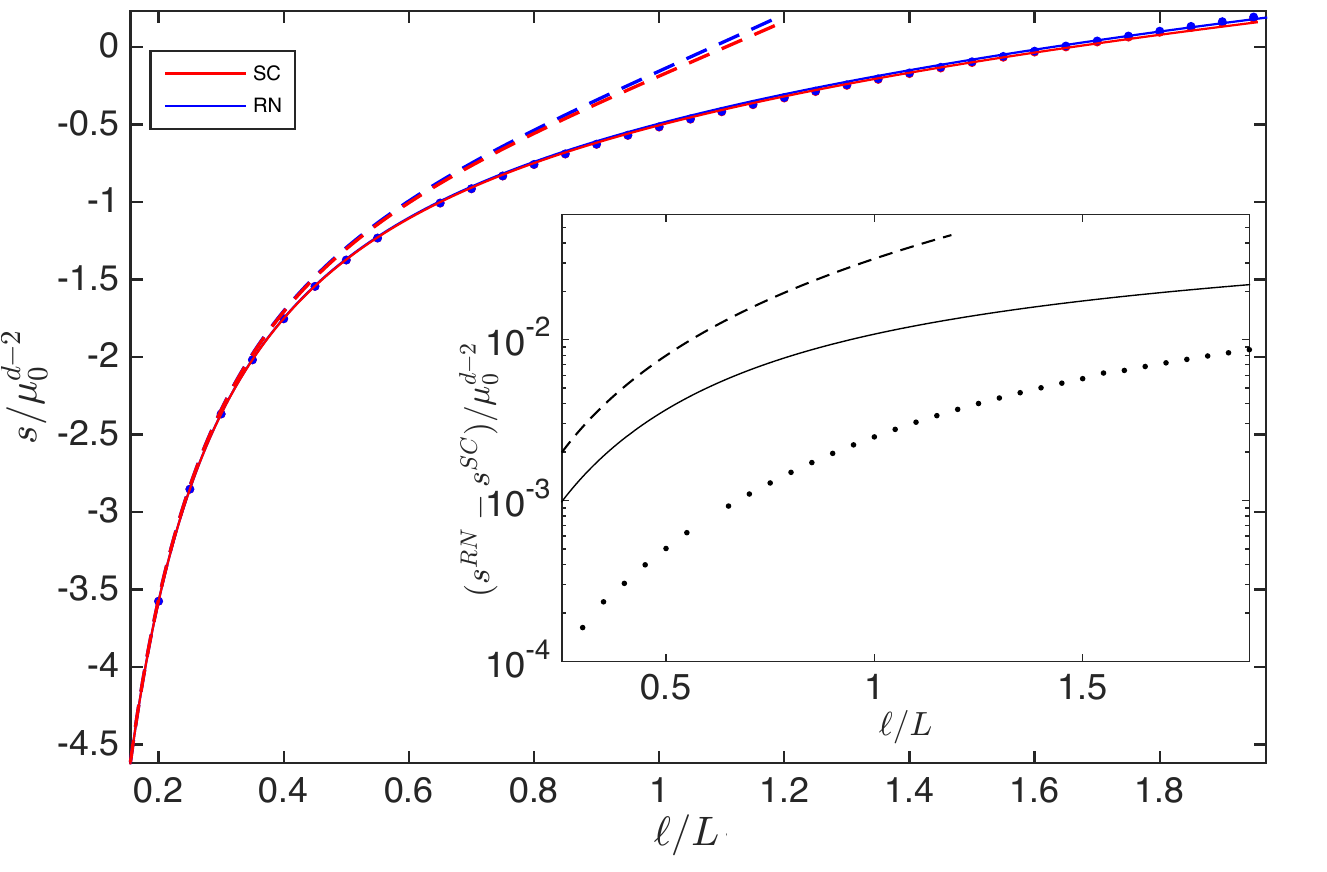}
\vspace{-1.2cm}
\caption{Finite part of the entanglement entropy, $s$, of a rectangular strip, $\tilde A$, of width $\ell$ in 2 spatial dimensions ($d=2+1$), where $s_{\tilde A}=s_{div}+s$ and $s_{div}$ contains the UV-cutoff. The blue dots are obtained by numerical  integration of eqs. (\ref{SAT}) and (\ref{lT}) in the Reissner-Nordstr\"{o}m  background with $f(z)=\frn$, $z_0=0.992$, $\mu=2.02$, $Q^2=\mu^2z_0^2\gamma^2$, $\gamma={d-2\over d-1}{L^4\over 2}$. Similarly, the red dots, hardly indistinguishable from the blue, correspond to the integration of these equations in the superconducting phase where $f(z)=\frn+\epsilon^2f_2$ and $f_2=f_2^a+f_2^b$ is given in eq. \ref{app:f}. We set $m^2L^2=0$, $\tilde z_0=1$, $q=4$ and $\epsilon=0.2$ corresponding to $T/T_c\sim 0.995$. The expansion parameter $\epsilon=\langle \OO/(2\Delta-d)\rangle^{1/\Delta}$ is defined in the same appendix. In the superconducting phase $\frn$ is given in terms of  $\mu_0=2.00$, eq. (\ref{app:mu0}), and $f_2^a$ in terms of $\kappa =-0.78$,  eq. (\ref{app:deltamu}). The solid lines are obtained from the analytical results for the superconducting and normal background from eqs. (\ref{lseries}) and (\ref{sseries}) by neglecting terms of $\OO(z_*/z_0)^{2d-1}$ and $\OO(z_*/z_0)^{2d-2}$, respectively. The dashed lines are obtained from the linear approximation, $z_*=\tilde z_* =\ell {d\ \Gamma\left({2d-1\over 2d-2}\right)\over 2\sqrt{\pi} \Gamma\left({3d-2\over 2d-2}\right)}$.  As anticipated, the analytical estimation of the entanglement entropy, calculated in the large $d$ limit, breaks down as $\ell$ increases. Nonetheless the qualitative behavior is very similar even for $d=2+1$ dimensions in the UV-boundary. Inset: difference between the finite parts of the entanglement entropies in both phases as a function of the strip length. As before, the dots correspond to the numerical results while the dashed and continuous lines are our analytical results corresponding to the linear approximation  and the subleading quartic correction, respectively. The lines are restricted to the region, in  $\ell$, where the estimations of the tip of the minimal surface in each background $\tilde z_*,\ z_*<z_0$, see figures  \ref{fig:lRN3d},\ref{fig:lH3d}.}
\label{pic:EE3d}
\end{figure}

Another relevant feature of the entanglement entropy eq.(\ref{ee_hairy_l}), that requires clarification, is that it does not obey the volume law.
It is well known that for a sufficiently large $\ell$, the finite contribution of the entanglement entropy at finite temperature satisfies the volume-law, not the area, i.e., a linear-in-$\ell$ term is expected. The analytical prediction eq.(\ref{ee_hairy_l}) does not reproduce such behavior since we are neglecting terms $(z_*/z_0)^{d}\ll 1$ in eqs.(\ref{lseries}),(\ref{sseries}). This is fine for small $\ell$ or, for a fixed $\ell$ and $T$, and a sufficiently large $d$.  However, for a large, but {\it fixed} $d$, the approximation breaks down for large $\ell$ since $z_*$ eventually becomes sufficiently close to the horizon $z_0$ so that $(z_*/z_0)^{d}\approx 1$. As seen in figures \ref{fig:lRN3d},\ref{fig:lH3d}, in principle a remedy to this problem is to include more terms in the expansions given in eqs. (\ref{lseries}), (\ref{sseries}). The first correction, proportional to $(z_*/z_0)^{d+1}$, coming from the $n=1$ term, still leads to an analytical, but cumbersome, expression for $z_*(\ell)$ in the case of $d=3$. Indeed, as is shown in figures \ref{fig:lRN3d},\ref{fig:lH3d}, by including this term, the analytical expression agrees with the numerical results up to larger values of $\ell$, however we do not yet observe the expected area law for and $\ell\to\infty$. Indeed, for any finite number of terms the approximation inevitably still breaks down at some finite $\ell$, and, already for $d=3$, the subleading correction, $\propto (z_*/z_0)^{2d-1}$, coming from the $n=1$ term in eq.(\ref{lseries}), leads to a fifth degree polynomial whose roots cannot in general be found numerically. Consequently we keep only the leading correction in the equations for $z_*$ and $\tilde z_*$ so our results are fully analytical. As is shown in figure \ref{pic:EE3d}, by including this additional term only, the analytical expression for the finite part of the entanglement entropy agrees well with the numerical results in the range of $\ell$ shown. However, as was mentioned previously, we do not yet observe the expected area law for $\ell \to \infty$. If, on the other hand, we carry out an analogous expansion in the parameter $1-(z_*/z_0)^{d}$, see appendix \ref{app:horizon}, we obtain the expected linear dependence of the entanglement entropy $s \propto \ell$. 

We also note that the finite subleading contribution, second term of eq.(\ref{ee_hairy_l}), that does not depend on the scalar, has already been reported in Ref. \cite{Ryu2006,Ryu2006a}. The dependence on the scalar, proportional to $\kappa<0$, is consistent with previous numerical results \cite{cliff2012}. It is smaller in the symmetry broken phase and its temperature dependence is not analytical at $T_c$ due to the different prefactors in the temperature dependence of the entanglement entropy in the broken and unbroken phase. We note the temperature enters both through $\tilde z_0$ and $\mu$ and it depends quadratically on the strip length $\ell$, for small $\ell$. These results illustrate the potential of $1/d$ expansions to obtain analytical results in problems where only numerical calculation were available.

%%%%%%%%%%%%%%%%%%%%%%%%%%%%%%%%%%%%
\section{Conclusions}
%%%%%%%%%%%%%%%%%%%%%%%%%%%%%%%%%%%%
We have studied the entanglement entropy and the conductivity in holographic superconductors at zero and finite temperature in the limit of large spatial dimensionality. The coherence peak of the conductivity becomes narrower and the ratio between the energy needed to break the condensate and the critical temperature decreases as the spatial dimensionality increases and have a well defined $d \to \infty$ limit. 
 This is a clear indication that the coupling of the scalar with the bulk is weaker in the large dimensionality limit. It would be interesting to explore whether there is a bound for these quantities in theories with gravity duals. We have computed the dependence on the dimensionality $d$ on the entanglement entropy at zero and close to the critical temperature and for the conductivity at zero temperature. Our results confirm the expectation that the entanglement entropy is smaller in the symmetry broken phase with a difference that decreases with the spatial dimensionality. These results are a strong indication that large $d$ expansions are a helpful tool to obtain analytical results in holography. 

\acknowledgments 
ARB thanks S. Gubser, C. Rosen, I. Takaaki, O. Dias, B. Gout\'eraux  for illuminating discussions. ARB acknowledges the support of the Department of Physics of the University of Cambridge. AMG was supported by EPSRC, grant No. EP/I004637/1.

\appendix

\section{Appendix: Entanglement entropy at $\boldsymbol{T\sim T_c}$}\label{App_EE}
 
The equations of motion, expressed in the coordinate $z=1/r$, are:
 \begin{equation}\label{eoms_finiteT}
\begin{split}
&\psi''-\left({\chi'\over 2}+{d-1\over z}-{f'\over f}\right)\psi'-\left({m^2L^2\over z^2f}-{q^2e^{\chi}\phi^2L^4\over f^2}\right)\psi=0,\\
& \phi''+\left({\chi'\over 2}-{d-3\over z}\right)\phi'-{2q^2L^2\psi^2\over z^2 f}\phi=0,\\
&\chi'={2z\over d-1 }\left(\psi'^2+{e^{\chi}q^2\phi^2\psi^2L^4\over f^2}\right)\hspace{-1mm},\\
& f'-{d\over z}f+{d\over z}={1\over (d-1)z}\left[ m^2L^2\psi^2+{z^4e^{\chi}\phi'^2L^4\over 2}+z^2f\left(\psi'^2+{q^2e^{\chi}\phi^2\psi^2L^4\over f^2}\right)\right].
\end{split}
\end{equation}
Close to the transition, the fields can be expanded in powers of $\ep\equiv\beta$, \cite{Herzog2010,Siopsis2012}, where $\psi(z\to0)\sim\beta({z\over z_0})^\Delta$, and $\Delta\equiv\Delta_+$ is the larger conformal dimension. More concretely,
\begin{equation}\label{expansions}
\psi\simeq \ep \psi_1+\ep^3\psi_3+\dots,\ \ \phi\simeq \phi_0+\ep^2\phi_2+\dots,\ \ f\simeq f_0+\ep^2f_2+\dots,\ \ \chi\simeq\ep^2\chi_2+\dots,
\end{equation}
For the purpose of the entanglement entropy calculation given in section \ref{sec:EE_Tc} we calculate analytically the first non-trivial terms of this field-expansion in the region where $\left(z\over z_0\right)^d\ll1$, which, for larger $d$ allows $z$ to approach $z_0$ with a better level of approximation than for small $d$. However, here we compute all the terms in the perturbative expansion up to $\OO (\epsilon^2)$. From the equations of $\phi$ and $f$ given in eq. (\ref{eoms_finiteT}), it is easy to see that the first zeroth order terms of these fields are:
\begin{equation}
\phi_0=\mu_0\left[1-\left({z\over z_0}\right)^{d-2}\right],\ \ f_0(z)=1-(1+Q^2) \left({z\over z_0}\right)^d+Q^2\left({z\over z_0}\right)^{2d-2},
\end{equation}
where $Q^2={\mu_0^2 z_0^2\gamma^{2}}$ and $\mu_0$ is the chemical potential at which the scalar field condenses. $\chi=0$ and $f=f_0$ corresponds to the Reissner-Nordstr\"{o}m black hole with planar topology. 
The equation for the first term in the expansion of $\psi$, eq. (\ref{expansions}), is well known:
\begin{equation}\label{eq:psi1}
0=\psi_1''-\left({{d-1\over z}-{f_0'\over f_0}}\right)\psi_1'+\psi_1\left({q^2L^4\phi_0^2\over f_0^2}-{m^2L^2\over z^2f_0}\right),
\end{equation}
giving the expected $(z/z0)^\Delta+\dots$ behavior close to the boundary. $\psi_1$ can be obtained by rewriting eq. (\ref{eq:psi1}) as a Sturm-Liouvillle eigenvalue problem, \cite{Siopsis2010,Pan2012}, and using as ansatz $\psi_1=z^\Delta F_0(z)$, $F_0=1-\alpha z^{d-1}$, $\alpha$ is given by the value, $\alpha_c$  that minimizes the following expression,
\begin{equation}\label{app:mu0}
M^2(\alpha)={\int_0^1dz\  z^{2\Delta-d+1}(1-z^d)\left[F_0'^2-\left({-m^2L^2\over 1-z^d}+{\Delta(\Delta-d)}-{\Delta d z^{d}\over 1-z^d}\right){F_0^2\over z^2}\right]\over\int_0^1dz\ z^{2\Delta-d+1}F_0^2q^2{(1-z^{d-2})^2\over 1-z^d}},
\end{equation}
and $\mu_0^2=M^2(\alpha_c)$.
The equation for $\chi_2$ is:
\begin{equation}
\chi'_2={2z\psi_1'^2\over d-1}+{2q^2zL^4\phi_0^2\psi_1^2\over(d-1)f_0^2}\equiv F_\chi(z),
\end{equation}
and thus $\chi_2(z)=\int_0^zdz'\ F_\chi(z')$.
Similarly, from the equation for $\phi_2$:
\begin{equation}\label{phi2eom}
\phi_2''-{d-3\over z}\phi_2'={2q^2\phi_0\psi_1^2\over z^2f_0}-{1\over2}\phi_0'\chi_2'\equiv F_\phi(z),
\end{equation}
and the leading behavior of $\phi_2$ is given by the homogeneous solution. Close to the horizon, $\phi_2$ is expected to receive corrections from the last two terms in eq. (\ref{phi2eom}). However, we impose such corrections, controlled by $\phi_0$ and $\psi_1'$, to satisfy the boundary condition $\phi_2(z_0)=0$ independently of the homogeneous solution. Therefore,   
\begin{equation}\label{app:phi2}
\begin{split}
&\hspace{4cm}\phi_2(z)=\phi_2^a(z)+\phi_2^b(z),\\
&\phi_2^a(z)\equiv\kappa\left[1-\left({z\over z_0}\right)^{d-2}\right],\ \phi_2^b(z)\equiv\int_z^{z_0}du\ u^{d-3}\int_u^{z_0}dv\ {F_\phi(v)\over v^{d-3}}.\\
\end{split}
\end{equation}
Finally, the equation for $f_2$ is given by:
\begin{equation}\label{eq:f2}
f_2'-{d\over z}f_2-{z^3 L^4 \phi_0' {\phi_2^a}' \over d-1}={1\over z(d-1)}\left(z^4L^4\phi_0'{\phi_2^b}'+z^2 f_0{\psi_1'^2}+{m^2L^2\psi_1^2}{q^2L^4 z^2\phi_0^2\psi_1^2\over f_0}\right)\equiv F_f(z).
\end{equation}
The previous equation can be integrated straightforwardly,
\begin{equation}\label{app:f}
\begin{split}
&\hspace{5cm}f_2(z)=f_2^a(z)+f_2^b(z),\\
&f_2^a(z)\equiv-{\mu_0\kappa  z_0^2L^4\over(d-1)(d-2)}\left[\left({z\over z_0}\right)^{d}-\left({z\over z_0}\right)^{2d-2}\right],\ \ f_2^b(z)\equiv-z^{d}\int_z^{z_0}\hspace{-2mm}du\ {F_f(u)\over u^d}\\
\end{split}
\end{equation}
In the large-$d$ limit, $f_2^a$ dominates over $f_2^b$. The only parameter to be determined is $\kappa $, which follows from the equation of $\psi_3$:
\begin{equation}
\begin{split}
&\hspace{2cm}\psi_3''-\left({{d-1\over z}-{f_0'\over f_0}}\right)\psi_3'+\left({q^2L^4\phi_0^2\over f_0^2}-{m^2L^2\over z^2f_0}\right)\psi_3=-{\cal T}\psi_0,\\
&{\cal T}\psi_0\equiv \left({f_2'\over f_0}-{\chi_2'\over2}-f_2{f_0'\over f_0^2}\right)\psi_0'+\left[{m^2L^2\over z^2f_0}f_2-{2q^2L^4\over f_0^3}(f_2\phi_0-f_0\phi_2)+{q^2\phi_0^2\chi_2L^4\over f_0^2}\right]\psi_0.\\
\end{split}
\end{equation}
From the previous equation and using eq. (\ref{eq:psi1}), it follows immediately, $0=\int_0^{z_0}dz\ {f_0\psi_0{\cal T}\psi_0\over z^{d-1}}$, which imposes a condition on $\kappa $:
\begin{equation}\label{app:deltamu}
\kappa ={\int_0^{z_0}\hspace{-1.5mm} dz\ {\psi_0'\psi_0\over z^{d-1}}\left({-{f_2^b}'}+{\chi_2' f_0\over2}+{f_2^bf_0'\over f_0}\right) \hspace{-1.3mm} +\hspace{-1.3mm} {\psi_0^2\over z^{d-1}}\left[{-m^2L^2f_2^b\over z^2f_0}+{2q^2L^4\over f_0}\left({f_2^b\phi_0^2\over f_0}-\phi_0\phi_2^b\right)-{q^2\phi_0^2\chi_2L^4\over f_0}\right] \over\int_0^{z_0}dz\  {\psi_0'\psi_0\over z^{d-1}}\left({{f_2^a}'}-f_2^a{f_0'\over f_0}\right) +{\psi_0^2\over z^{d-1}}\left[{m^2L^2f_2^a\over z^2f_0}-{2q^2L^4\over f_0}\left(f_2^a{\phi_0^2\over f_0}-\phi_0\phi_2^a\right)\right]}.
\end{equation}

Finally, from eq. (\ref{app:phi2}), for some $\epsilon>0$, the chemical potential is given by $\mu\sim \mu_0+\epsilon^2(\kappa +\phi_2^b(0))+\OO(\epsilon^4)$, $\rho=\mu_0+\epsilon^2\kappa $ and the temperature $T<T_c$:
\begin{equation}\label{app:T}
\begin{split}
&\hspace{2.5cm}{T\over\rho^{1\over d-1} }=-{f'(z_0)e^{-\chi(z_0)/2}\over 4\pi\left(\mu_0+\epsilon^2\kappa \right)^{1\over d-1}}\sim\\
&\sim{-f_0'(z_0)\over 4\pi\mu_0^{1\over d-1}}\left[1+\epsilon^2\left({f_2'(z_0)\over f_0'(z_0)}-{\chi_2(z_0)\over 2}-{\kappa \over (d-1)\mu_0f_0'(z_0)}\right)\right],\\
\end{split}
\end{equation}
where, $f_0'(z_0)=-d+\mu_0^2{(d-2)^2\over 2d-2}$. For $\epsilon=0$, the previous equation gives an estimation for the critical temperature. This expression is more complicated that the one given in Ref.\cite{Emparan2014}, in which the backreaction of the scalar on the geometry is neglected. Notice however, that, we were not after an alternate result for $T_c$, in fact, in this section we have not used the large-$d$ limit since we explicitly look for all the terms that modify the geometry close to the phase transition. In order to analytically evaluate the leading correction on the entanglement entropy of a strip with its complement, section \ref{sec:EE_Tc}, we take the leading correction on the blackening function,  $f_2^a(z)$, given in eq. (\ref{app:f}).

\section{Large $\boldsymbol \ell$ limit of the entanglement entropy at finite temperature and fixed $\boldsymbol d$}\label{app:horizon}
In the RN background, from eq. (\ref{lT}),
\begin{equation}
{\ell\over2}=\int_0^{z_*}dz{z^{d-1}\over\sqrt{f(z)\left(z_*^{2d-2}-z^{2d-2}\right)}}.
\end{equation}
Let us split the region of integration into two: $[0,z_*]=[0,z_a]\cup(z_a,z_*]$, for some $0<z_a<z_*$ and let us rename the integral in the first interval as $\ell_1/2$. In the second interval we change variables to $z=z_0-\epsilon$ and expand the integrand for $\epsilon/z_0\ll 1 $.
\begin{equation}
\begin{split}
{\ell\over2}&\sim{\ell_1\over 2}-\int_{z_0}^{z_0-z_*}d\epsilon{{z_0^{d-1/2}\over\sqrt{(d-2)Q^2-d}\sqrt{z_0^{2d-2}-z_*^{2d-2}}\sqrt{\epsilon}}}+\dots\sim\\
&\sim\ell_1+{z_0^{d-1/2}\over2\sqrt{(d-2)Q^2-d}}\left[-\left(z_0-z_*\over z_0^{2d-2}-z_*^{2d-2}\right)^{1/2}+{z_0-z_a\over\sqrt{z_0^{2d-2}-z_*^{2d-2}}}\right]+\dots.\\
\end{split}
\end{equation}
In spite of the explicit dependence of the third term on $z_*$, in the limit $z_*\to z_0$ we can take it as a divergent contribution, $\ell_{div}$, while the middle term remains finite, thus:
\begin{equation}\label{app:term}
-\left(z_0-z_*\over z_0^{2d-2}-z_*^{2d-2}\right)^{1/2}\sim(\ell-\ell_1){\sqrt{(d-2)Q^2-d}\over z_0^{d-1/2}}-\ell_{div}.
\end{equation}
Similarly, splitting the integral of the entanglement entropy, eq. (\ref{SAT}), into the same regions the integrations carries analogously,
\begin{equation}
s_{\tilde A}\sim s_1-{\sqrt{z_0}\over \sqrt{(d-2)Q^2-d}}\left(z_0-z_*\over z_0^{2d-2}-z_*^{2d-2}\right)^{1/2}+{\sqrt{z_0}\over \sqrt{(d-2)Q^2-d}}\left(z_0-z_a\over z_0^{2d-2}-z_*^{2d-2}\right)^{1/2},
\end{equation}
where $s_1$ contains the UV-cutoff and the last term is also divergent in the limit $z_*\to z_0$. The middle term can be substituted using eq. (\ref{app:term}) which leaves a term proportional to $\ell$. In the limit $z_*\to z_0$, $\ell\to\infty$ however, the term in eq. (\ref{app:term}) is regularized by $\ell_{div}$, yielding a finite term.

\section{Electrical conductivity at $\boldsymbol{T>0}$}\label{App:cond}
The boundary conditions near $z\to0$ are:
\begin{equation}\label{Ax:d}
\tilde A_x(z,\omega)=
\begin{cases}
A_x^{(0)}+A_x^{(1)}z,& \ d=3 \\
A_x^{(0)}+A_x^{(1)}z^2+{A_x^{(0)}\omega^2\over2}{z^2\log{\Lambda\over z}}, & \ d=4 \\
\Axz + \Axo \sqrt{
    2\over \pi \omega} \left[-z \cos(z \omega) + {1\over \omega}\sin({z  \omega})\right], & \ d=5 \\
\Axz + \Axo \sqrt{
    2\over \pi \omega} \left[-{3z\over \omega} \cos(z \omega) - {z^2}\sin({z \omega})+{3\over \omega^2}\sin(z \omega)\right], & \ d=7 \\
\Axz + \Axo \sqrt{
    2\over \pi \omega} \left[\left(z^3-{15z\over \omega^2}\right) \cos(z \omega)+\left({15\over \omega^3} - {6z^2\over\omega}\right)\sin({z \omega})\right], & \ d=9 \\
\end{cases}
\end{equation}
$\Lambda$ is a cutoff which affects only the imaginary part of $\sigma$. We take $\Lambda=1$. From the above expressions and the eq. (\ref{conductivity}), the conductivity is:
\begin{equation}\label{cond:d}
\sigma=
\begin{cases}
{A_x^{(1)}\over i\omega A_x^{(0)}}, & \ d=3 \\
{2A_x^{(1)}\over i\omega A_x^{(0)}}+{i\omega\over 2}, & \ d=4 \\
{A_x^{(1)}\over i A_x^{(0)}}{\sqrt{2\omega\over\pi}}, & \ d=5 \\
{3A_x^{(1)}\over i A_x^{(0)}}\omega^{3/2}\sqrt{2\over \pi}, & \ d=7 \\
{15A_x^{(1)}\over i A_x^{(0)}}\omega^{5/2}\sqrt{2\over \pi}, & \ d=9 \\
\end{cases}
\end{equation}

\newpage

\bibliographystyle{my_style_AROMERO}
\bibliography{largeDbib}

\end{document}